\documentclass[a4paper,11pt]{article}
\pdfoutput=1 
\usepackage{jheppub} 
\usepackage{inputenc}
\usepackage[makeroom]{cancel}
\usepackage[normalem]{ulem}
\usepackage[usenames,dvipsnames]{xcolor}
\usepackage{slashed,soul,multirow, makecell}
\newcommand{\cmmnt}[1]{\ignorespaces}

\title{Jet substructure probe to unfold singlet-doublet dark matter in the presence of non-standard cosmology}

\author[a]{Prasanta Kumar Das,}
\author[b]{Partha Konar,}
\author[a]{Saumyen Kundu,}
\author[~b]{and Sudipta Show}

\affiliation[a]{Department of Physics, Birla Institute of Technology and Science-Pilani, K. K. Birla Goa campus, NH-17B, Zuarinagar, Goa 403726, India}
\affiliation[b]{Theoretical Physics Division, Physical Research Laboratory, Shree Pannalal Patel Marg, Ahmedabad 380009, Gujarat, India}

\emailAdd{pdas@goa.bits-pilani.ac.in}
\emailAdd{konar@prl.res.in}
\emailAdd{p20170022@goa.bits-pilani.ac.in}
\emailAdd{sudipta@prl.res.in}

\abstract{ We examine the singlet-doublet fermionic dark matter model, where the non-thermal production of the dark matter in light of a non-standard cosmology demands a significantly large interaction rate than the typical radiation-dominated Universe.
Despite being a model of freeze-in light dark matter and heavy mediator, the characteristic long-lived particle searches at the collider experiment and the displaced vertex signature do not help in probing such a dark sector since this non-standard interaction mandates nearly prompt decay. 
We make a counterproposal to probe such signal with di-fat-jets generated from the boosted decays of massive vector bosons and Standard Model Higgs, along with the substantial missing transverse momentum to probe the dark matter at LHC. 
Interestingly, substructure variables associated with these fat jets have an additional handle to tackle the extensive QCD background as it encodes implicit footmarks of their origin. We adopt the multivariate analysis with the booted decision tree to constrain the measured relic density allowed parameter space of dark matter in the presence of the modified cosmological scenario. Our study shows how the non-trivial expansion affects dark matter production in the early Universe and alters the required search strategies at colliders. This probe provides the best discovery prospect at the HL-LHC for extended parameter space now opened up in the dark sector.
}
\preprint{\today}

\begin{document} 

\maketitle

\flushbottom

%
\section{Introduction}
Numerous indirect pieces of evidence at diverse length scales of our Universe have conclusively established the omnipresence of some non-luminous, non-baryonic mysterious matter commonly known as dark matter (DM). It is interesting that so far, all such clues that are based on different celestial observations, such as the flatness of the galaxy rotation curve~\cite{Sofue:2000jx}, the studies of bullet cluster collision~\cite{Clowe:2006eq}, gravitational lensing of distant galaxies where only the gravitational interaction of DM is involved. Another indirect hint comes from experiments like WMAP~\cite{Hinshaw_2013} and PLANCK~\cite{Planck:2018vyg}, probing the anisotropy of the cosmic microwave background, which quantifies the present amount of dark matter of the Universe with great precision. 
Despite all these, the primary quest on the particle nature of dark matter, its interactions with different SM or other dark-sector particles and the production mechanism at an early stage of the Universe still remain elusive. Based on these observed properties, there exists a plethora of models for DM. Among all such variants, the weakly interacting massive particle (WIMP) paradigm~\cite{Green:2002ht, Chang:2017gla,Chang:2017dvm,Visinelli:2017qga,Arcadi:2017wqi,Choubey:2017yyn,Reinert:2017aga,Evans:2017kti,Garny:2018icg,Blanco:2019hah,Bhardwaj:2018lma, Bhardwaj:2019mts, Konar:2020wvl, Konar:2020vuu, Heurtier:2019beu,Habermehl:2020njb,Xing:2021pkb,Borah:2022byb,Belanger:2022qxt, Gines:2022qzy,Bernal:2022wck,Kundu:2021cmo,Medina:2021ram,Tallman:2022nts,Kang:2022zqv,Dutta:2022wdi} is the most explored one till date. Non-observation of any DM signature so far in direct detection~\cite{Akerib:2016vxi, Zhang:2018xdp, Aprile:2018dbl} and indirect detection~\cite{MAGIC:2016xys} experiments as well as in the collider frontiers~\cite{Chatrchyan:2012xdj, Aad:2012tfa} severely constrain this scenario.
Null results, so far, from all these experiments, especially the stringent constraints from direct detection, put a question mark on the scale of the DM interaction and encourage us to examine alternative paradigms of DM production mechanism. Non-thermal origin of DM is one such exciting scenario that naturally explains the null results thanks to its feeble interaction with the visible sector. Due to this same rationale, the DM never gets thermalized with the primordial bath in the early Universe; it is instead produced gradually from the scattering and/or the decay of visible particles in the early Universe until the production mechanism freezes in. Recently the feebly interacting massive particle (FIMP)~\cite{Hall:2009bx, Konar:2021oye, Ghosh:2021wrk, Chakrabarty:2022bcn} paradigm of DM has also got significant attention in the literature.

Since dark matter production is the phenomenon of the early Universe, the history of the early Universe is crucial. Typically, most DM studies consider the radiation-dominated early Universe, although there is no direct evidence to date about the energy content of the Universe at higher temperatures. So, it is prudent to take a holistic view considering the possibility of domination in species other than radiation in the energy budget~\cite{DEramo:2017gpl, DEramo:2017ecx} during the pre-Big Bang Nucleosynthesis (BBN) era. In this work, we adopt the scenario of fast expansion where the species ($\phi$), red-shifting faster than radiation, dominates the energy content of the early Universe.

In this work, we focus on the freeze-in production of dark matter where direct and indirect searches can hardly provide any constraint due to the feeble interaction of DM with the particles of the visible sector. One cannot directly produce such DM at the collider too. However, exploring the production of other dark sector particles~\footnote{Freeze-in production requires additional dark sector particle(s) with enough interaction with the SM to stay at thermal equilibrium and produce DM from decay.} in high-energy collision events can provide important inputs on this sector and DM. Due to this weaker interaction strength, interesting constraints can emerge, especially in the parameter space where the decay of the long-lived particle to DM with an SM partner.
Thus, a viable freeze-in dark matter scenario can be characterized by signatures like disappearing tracks and displaced vertex at LHC if the decay process is localized within a specific range of the detector. This requires DM mass of the order of keV~\cite{Hall:2009bx, Calibbi:2018fqf, No:2019gvl}. Interestingly, this is also a typical mass scale that astrophysical experiments can probe in the context of warm dark matter~\cite{Viel:2013fqw, Yeche:2017upn, Irsic:2017ixq}. Such warmness suppresses the structure formation at a short distance, which solves the small-scale problem in the $\Lambda$CDM (the standard picture of cosmology)~\cite{Bode:2000gq, Lovell_2012}.
Authors in the refs~\cite{Calibbi:2018fqf, No:2019gvl} have studied the frozen-in warm dark matter and demonstrated the connection between the long-lived particles along with the displaced vertex signature as a powerful tool in the context of singlet doublet dark matter scenarios. 

In this work, we consider the singlet-doublet model~\cite{Yaguna:2015mva, Fiaschi:2018rky, Restrepo:2019soi, Arcadi:2018pfo, Esch:2018ccs, Calibbi:2018fqf, Maru:2017pwl, Maru:2017otg, Xiang:2017yfs, Abe:2017glm, Banerjee:2016hsk, Horiuchi:2016tqw, Calibbi:2015nha, Cheung:2013dua, Cohen:2011ec, Enberg:2007rp, DEramo:2007anh, No:2019gvl, Barman:2019aku,  DuttaBanik:2018emv, Barman:2019tuo, Bhattacharya:2018fus, Bhattacharya:2015qpa, Bhattacharya:2017sml, Restrepo:2015ura, Freitas:2015hsa, Cynolter:2015sua, Bhattacharya:2016lts, Bhattacharya:2016rqj, Wang:2018lhk, Abe:2019wku, Barman:2019oda, Konar:2020wvl, Konar:2020vuu, Konar:2021oye} to demonstrate the impact of non-standard cosmology both in the dark matter phenomenology as well as in the prospect of collider probes. We exhibit that the introduction of modified cosmology demands a larger interaction strength than the usual (standard cosmological) scenario~\cite{Elor:2021swj,Bhattiprolu:2022sdd}, and naturally, it also alters the probing methodology at the collider. A larger interaction rate demands prompt decays rather than giving a displaced vertex signature. We suggest probing this final state with the production of $Z/W/h$-bosons, along with a substantial missing transverse momentum from DM particles at the LHC. The bosons are also expected to be significantly \emph{boosted} due to their production from heavy mediators. Identities of their individual decay products are mostly undetectable, being submerged in localized pixels at the detector along the boosted direction, and such detections are expected to have poor efficiency. On the other hand, directly looking at total hadronic decay products in a large radius jet can be highly advantageous, especially if one looks at the formation of substructures inside such jets to assess their origin. Substructure variables can significantly improve the detection of such events compared to  QCD jet background events.
We look at the jet substructure of the boosted $Z/W/h$-jets coupled with sophisticated multivariate analysis (MVA) to probe the significant part of the parameter space, inaccessible with the displaced vertex searches of this model.

This paper is organized as follows. In \autoref{model}, we sketch the singlet-doublet dark matter model discussing different decay modes of the heavier fermions, which have an essential role both in frozen-in DM production and at the LHC search. In \autoref{NScosmo}, we briefly review the non-standard cosmology that arises due to the existence of another species in the early Universe. In the next section (\autoref{DMpheno}), we discuss the phenomenology of the frozen-in DM in the context of non-standard cosmology. In \autoref{collider}, we addressed the collider setup in detail and the search strategy using the substructure variables. We also include the multivariate analysis in this section before presenting our main results. Finally, we summarise and conclude in \autoref{conclusion}.

%
\section{Structure of the model}\label{model}
We execute our analysis in the singlet-doublet model~\cite{No:2019gvl, Konar:2020wvl, Konar:2020vuu, Konar:2021oye}, which consists of  a hyperchargeless singlet Dirac fermion ($\chi$) and an $SU(2)_L$ Dirac fermion doublet ($\Psi$) 
\begin{equation*}
\Psi = 
\begin{pmatrix}
\psi^+ \\
\psi^0\end{pmatrix}
\end{equation*}
with hypercharge $1/2$.
Here, we employ the discrete $\mathcal{Z}_2$ symmetry under which the BSM fields change sign while all the SM fields are assigned with even $\mathcal{Z}_2$ charges. The Lagrangian for the BSM fermions can be written as
\begin{equation}
\mathcal{L}_f=i\overline{\chi}\gamma^\mu\partial_\mu\chi+i\overline{\Psi}\gamma^\mu D_\mu\Psi+m_\chi\overline{\chi}\chi+m_\Psi\overline{\Psi}\Psi+y\overline{\Psi} H\chi + h.c.
\label{eq:lag_f}
\end{equation}
Both the neutral particles $\psi^0$ and $\chi$ get mixed after the EW symmetry breaking because of the Yukawa type interaction term in Eq.~\eqref{eq:lag_f}. After mixing, the new mass eigenstates are $\chi_1$ and $\chi_2$ having masses $m_1$ and $m_2$, respectively. We make simple assumptions $m_2\gg m_1$ and $y\ll 1$, which are needed for the freeze in production to yield a correct amount of dark matter relic density. In this limit, one gets $m_1\approx m_\chi$, $m_2\approx m_\Psi$. Hence, the simplified form of the singlet-doublet mixing angle becomes
\begin{equation}
    \sin\theta\approx\frac{y v}{\sqrt{2}(m_2-m_1)},
\end{equation}
where $v$ is the vacuum expectation value of the SM Higgs. A radiative mass splitting between the charge states ($\psi^\pm$) and neutral state ($\psi^0$) is generated due to the loops of the EW gauge boson~\cite{Thomas:1998wy, Cirelli:2009uv}, so that mass splitting $\mathcal{O} [260-340]$ MeV can be produced for the considered range of $m_\Psi \in$ $[100-2000]$ GeV. For simplicity, we would neglect the mass splitting for the dark matter phenomenology and assume that the masses of all the heavy states are equal such that $m_{\psi^{\pm}} = m_2 = m_\Psi$. The interactions of the dark matter after the EW symmetry breakings are $h-\chi_2-\chi_1$, $Z-\chi_2-\chi_1$, $W^\pm-\psi^\mp-\chi_1$ where the first one responsible for the Yukawa interaction expressed in Eq.~\eqref{eq:lag_f} and rest of the interactions come from the singlet doublet mixing originated by the Yukawa interaction after the symmetry breaking. Corresponding decays widths for these  heavy BSM states can be given by,
\begin{align}
    &\Gamma(\chi_2\to h\chi_1)=\frac{y^2}{32\pi m_\Psi^2}\left[(m_\Psi+m_\chi)^2-m_h^2\right]\lambda(m_\Psi,m_\chi,m_h), \label{eq:decay_width1} \\
    &\Gamma(\chi_2\to Z\chi_1)=\frac{y^2}{32\pi m_\Psi^2}\frac{\left[(m_\Psi-m_\chi)^2-m_Z^2\right]\left[(m_\Psi+m_\chi)^2+2m_Z^2\right]}{m_\Psi^3(m_\Psi-m_\chi)^2}\lambda(m_\Psi,m_\chi,m_Z), \label{eq:decay_width2} \\ 
    &\Gamma(\psi^\pm\to W^\mp\chi_1)=\frac{y^2}{32\pi m_\Psi^2}\frac{\left[(m_\Psi-m_\chi)^2-m_W^2\right]\left[(m_\Psi+m_\chi)^2+2m_W^2\right]}{m_\Psi^3(m_\Psi-m_\chi)^2}\lambda(m_\Psi,m_\chi,m_W)
       \label{eq:decay_width3}
\end{align}
where, $\lambda(a,b,c)=\sqrt{a^4+b^4+c^4-2a^2b^2-2b^2c^2-2c^2a^2}$. 

%
\section{A framework with fast expanding Universe}\label{NScosmo}

The standard model of cosmology assumes that the energy budget of the early Universe was radiation-dominated prior to the Big Bang Nucleosynthesis (BBN) era. Although, the lack of any observational evidence can not rule out the possibility that some other field could have dictated the evolution of the Universe rather than radiation, where the corresponding field redshifts smaller or larger than the radiation. Many previous studies~\cite{DEramo:2017gpl, DEramo:2017ecx, Konar:2020vuu, Konar:2021oye,Haque:2023yra} have demonstrated the non-trivial impact of such modified cosmology on DM phenomenology.

The non-standard cosmology can be simply realized by considering the contribution of the extra new species ($\phi$) along with the radiation component in the energy content of the Universe. Here, we consider the non-standard scenario where the parameter of the equation of state ($\omega$) is larger than that of the radiation ($\omega_{rad}$=1/3). One can parameterize the same by writing the corresponding energy density as $\rho_\phi\propto a^{-3(1+\omega)}$, which can be re-expressed as $\rho_\phi\propto a^{-(4+n)}$ with $\omega=\frac{1}{3}(n+1)$ provided $n>0$. Hence, the modified version of the Hubble rate~\cite{DEramo:2017gpl, DEramo:2017ecx, Konar:2020vuu, Konar:2021oye} of the Universe can be delineated as
\begin{align}
	H^2=\frac{\rho_{rad}+\rho_\phi}{3 M_P^2},
\end{align}
where $M_P$ represents the reduced Planck mass and the $\rho_{rad}$ and $\rho_\phi$ stand for the radiation energy density and the energy density of the species ($\phi$), respectively. The total energy density as a function of temperature can be expressed as
\begin{align}\label{eq:totalrho}
	\rho(T) &= \rho_{\rm rad}(T)+\rho_{\phi}(T)\\
	&=\rho_{\rm rad}(T)\left[1+\frac{g_* (T_r)}{g_* (T)}\left(\frac{g_{*s}(T)}{g_{*s}(T_r)}\right)^{(4+n)/3}\left(\frac{T}{T_r}\right)^n\right],
\end{align}
with $\rho_{rad}=\frac{\pi^2}{30}g_*(T)T^4$, $g_*$ and $g_{*s}$ denote the relativistic degrees of freedom for energy and entropy, respectively. Here, $T_r$ represents the temperature at the end of the fast-expanding Universe era, after which the Universe enters into a radiation-dominated phase. Thus the non-standard behaviour of the early Universe can be described simply by the set of two parameters $(n, T_r)$. A larger value of $n$ or a smaller value of $T_r$ entails that the Universe is red shifting faster. It is important to note that one can get back the standard picture in the absence of the extra species ($\phi$) with energy density, $\rho=\rho_{\text{rad}}$. The BBN constraints on the number of relativistic degrees of freedom put a lower bound on $T_r(\ge (15.4)^{1/n}$ MeV$)$. An interesting scenario, $n=2$ ($\omega=1$), is very well accepted as the kination domination~\cite{DEramo:2017gpl, DEramo:2017ecx, Konar:2020vuu, Konar:2021oye}, which we consider our case study throughout our whole analysis. In the next section, we discuss the effects of the modified cosmology on the DM production mechanism.

%
\section{Dark Matter Phenomenology}\label{DMpheno}

In our setup, the doublet ($\Psi$) becomes part of the thermal bath because of its gauge interaction, while the dark matter ($\chi$) never gets thermalized with the bath due to the feeble Yukawa coupling ($y\ll 1$). In the early Universe, the dark matter is produced gradually via freeze-in from the scattering and the decay of the bath particles starting from zero (or negligibly small) number density. Since the DM production crucially depends on the early history of the Universe, the freeze-in number density of DM is built upon the Yukawa coupling ($y$) along with the non-standard parameters $n$ and $T_r$.

\begin{figure}[tb!]
	\centering
	\includegraphics[width=0.9\textwidth]{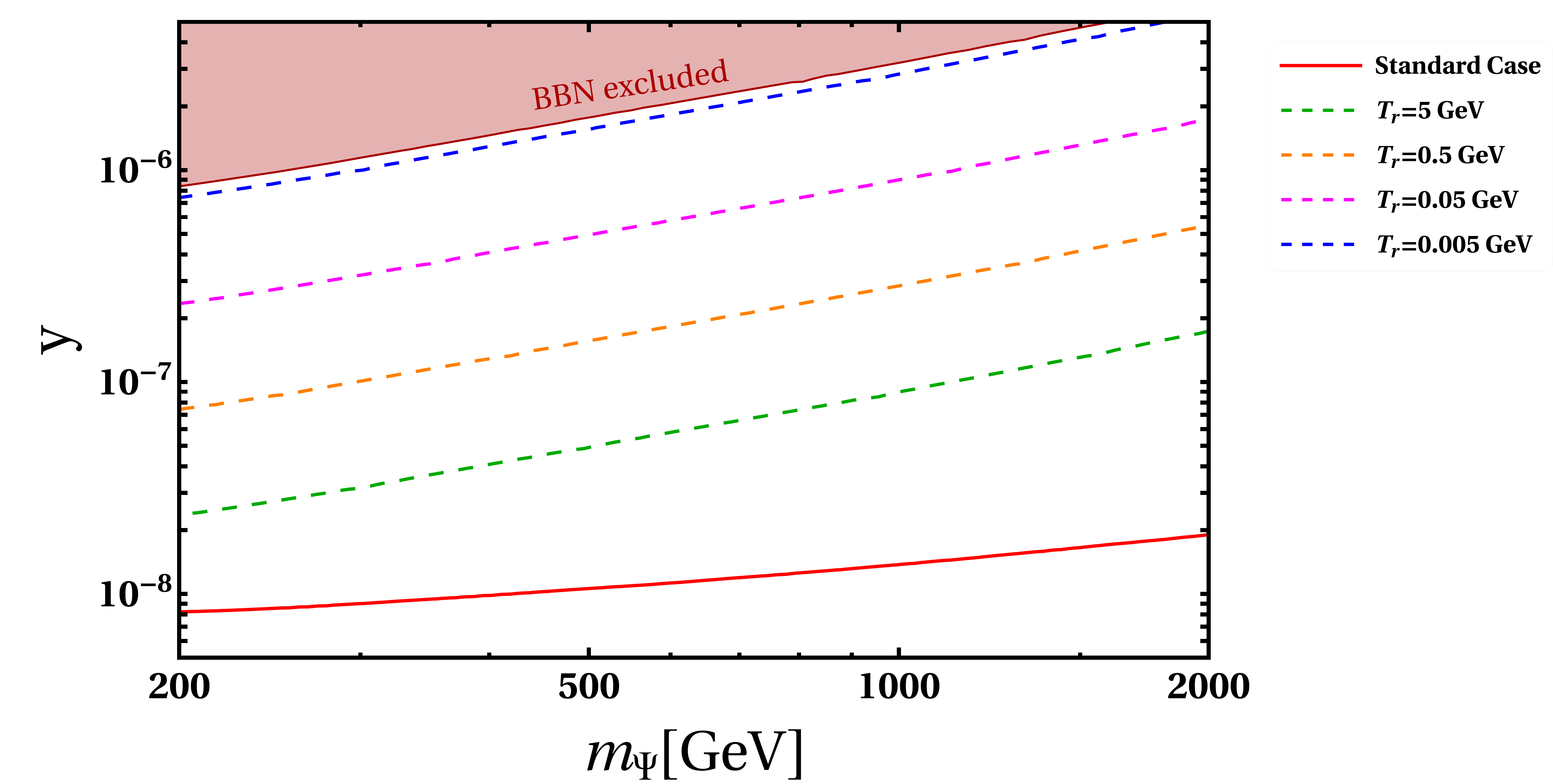} 
	\caption{ 
		Contours of the measured dark matter relic density $\Omega h^2 = 0.12$ are shown for different non-standard cosmology parameters ($T_r$) with a fixed value of $n=2$ on the Yukawa coupling (y) and doublet mass ($m_\psi$) plane. Non-standard cases are plotted with dashed lines for corresponding $T_r$ values, which opened up a vast region of parameter space over and above the standard cosmology ($n=0 ~or~ T_r \to \infty$) confined on a solid line in this plot. Dark matter mass is also fixed at  $m_{\chi} = 12$ keV.}
	\label{fig:DMrelic}
\end{figure}

Here, the dark matter production takes place in two steps: before and after the \emph{Electroweak Symmetry breaking} (EWSB). \emph{Before} EWSB, the dark matter production is carried out by different scatterings ($HH\to \chi\chi$, $\Psi\Psi\to \chi\chi$) and the decay ($\Psi\to H\chi$) processes. Since the scattering is proportional to $y^4$  while the decay varies as $y^2$, one can safely neglect the scattering contributions in the limit $y\ll 1$. \emph{After} EWSB, some additional channels open up due to the singlet doublet mixing. Here, the processes contributing to the DM productions are the scatterings ($SM SM \to \chi_1 \chi_1$, $SM SM \to \chi_2 \chi_1$) with both the processes mediated by the SM Higgs and $SU(2)_L$ gauge bosons and decays ($\chi_2 \to h(Z)\chi_1$, $\psi^\pm\to W^\pm \chi_1$). If $y\ll 1$, once again, all the scattering processes remain subdominant and would have a negligible impact compared to decays in producing DM. Since decays significantly dominate over the scattering in both regimes, it is very important to define the thermally average decay width, 
\begin{equation}
\langle\Gamma_\Psi\rangle=\langle\Gamma_{\Psi\to H \chi}\rangle~\Theta(T-T_{\rm EW})+\langle\Gamma_{\chi_2\to h \chi}+\Gamma_{\chi_2\to h \chi}+\Gamma_{\psi^\pm\to h \chi}\rangle~\Theta(T_{\rm EW}-T)
\end{equation} 
with $T_{EW}(=$160 GeV$)$ is the temperature of the EWSB. Now, one can track the dark matter abundance by solving the following Boltzmann equation,
 \begin{align}
	&\frac{dY_{\chi}}{dz}=
	\frac{ {\langle\Gamma_{\Psi}\rangle}}{Hz} Y_\Psi^{\rm eq}(T),
	\label{eq:Boltz1} 
\end{align} 
where $Y_i$ represents the abundance of the $i^{\text{th}}$ species, $z(=m_\chi/T)$ is a dimensionless variable and ${\langle\Gamma_{\Psi}\rangle}(=\Gamma_{\Psi}\frac{K_1(x)}{K_2(x)})$ refers to the thermally averaged decay rate. One can obtain the relic density of the dark matter by using the following relation
\begin{align}
	\Omega_\chi h^2=2.755\times 10^{8}\times m_\chi\times Y_\chi({x\to\infty})
\end{align}
In \autoref{fig:DMrelic}, we have shown the contour of the correct relic density of the dark matter in the plane of the Yukawa coupling and doublet mass. When the value of the parameter ($T_r$) reduces, then one needs larger Yukawa coupling ($y$) for the fixed values of ($n$), and dark matter mass to satisfy the relic density constraints as the effects of the fast expansion lasts for a longer time. Hence, in this modified scenario, the entire parameter space above the standard case becomes allowed for different values of $T_r$ (and $n$).
It is important to mention that any large Yukawa coupling is not allowed for our considered mass range of the heavy fermion since there is an upper limit on $T_r$ from BBN. We have depicted the BBN excluded region by the shaded red area in the upper portion of the plot.
\begin{table}[t] 
	\small
	\centering
	\begin{tabular}{|l|c|c|c|c|c|c|c|c|}
		\hline
		Benchmark & $y$ 	& $m_\psi$ 	&  $m_\chi$ 	& $n$ &  $T_r$  & $\Omega_\chi h^2$ 	& $\Gamma_\Psi$  & $c\tau_\Psi$\\     
		Points         &  	& [GeV] 		&  [KeV] 		&  	  &  [GeV]	  	 & 		&		[GeV]    & [cm]	\\     \hline  \hline
		BP-1 & $1\times10^{-6}$ & 860 		&  12  		& 2 	  &    	0.031	 &  	0.12	&	$2.53\times10^{-11}$	& 0.0008	\\     \hline
		BP-2 & 1$\times10^{-7}$ & 810 		&  12  		& 2 	  &    2.8		 &  		0.12	&$2.38\times10^{-13}$	& 0.083	\\     \hline
		BP-3 & $1.275\times10^{-8}$ & 825 		&  12  		& 0 	  &    	-	 &  	0.12	& $3.94\times10^{-15}$	&	5.025	\\     \hline
	\end{tabular}
	\caption{Benchmark Points considered in the analysis. }
	\label{tab:BP}
\end{table}

Before moving further to describe the possible probe at the LHC, let us construct relevant benchmarks for our current study. We demonstrate our results considering the benchmark points in \autoref{tab:BP}, which satisfy all the dark matter constraints. The first two benchmark points are relevant for non-standard cosmology, while the last one (BP-3) is placed overlapping the standard case for presentation purposes. All these points satisfy the relic density constraint for suitable values in $T_r$ (and $n$).
It is to be noted that the decay length for BP-3 is around 5 cm, so this can be probed better through the displaced vertex (DV) signature as the length of the tracker is around 1cm. Although, for BP-1 and BP-2, the decay length is sub-centimetre since the presence of alternative cosmology dictates very rapid decay. As a result, the DV signature can not be a better tool to probe our interesting region of parameter space; rather, the study of jet substructure plays a pivotal role in collider searches.

\section{Probing dark sector at the LHC}\label{collider}

This section discusses the possible collider signature to probe the model. As mentioned already, tiny coupling makes direct dark matter production at the LHC difficult. Still, the presence of other dark sector particles with relatively significant coupling makes a case for their pair production which eventually decays to produce the DM candidate at the collider.
The conventional and popular search strategy of FIMPs at colliders has been to look for the displaced signatures of the mediators owing to their large decay length. Other search channels, such as disappearing tracks or mono-jet search, are less sensitive at LHC in the standard cosmological scenarios. Now, with a larger interaction rate as necessitated by the alternative cosmological scenario discussed in \autoref{NScosmo}, the displaced vertex search loses its significance, and prompt searches take over. In such a scenario, decay of the heavy mediators $\Psi^\pm$ and $\chi_2$ exert a substantial boost on the decay products leading to the emission of large radius hadronic clusters. Here we intend to examine the possible substructure of these clusters in the reconstructed final states of the events. 

\begin{figure}[!t]
	\centering
	\includegraphics[width=0.495\linewidth]{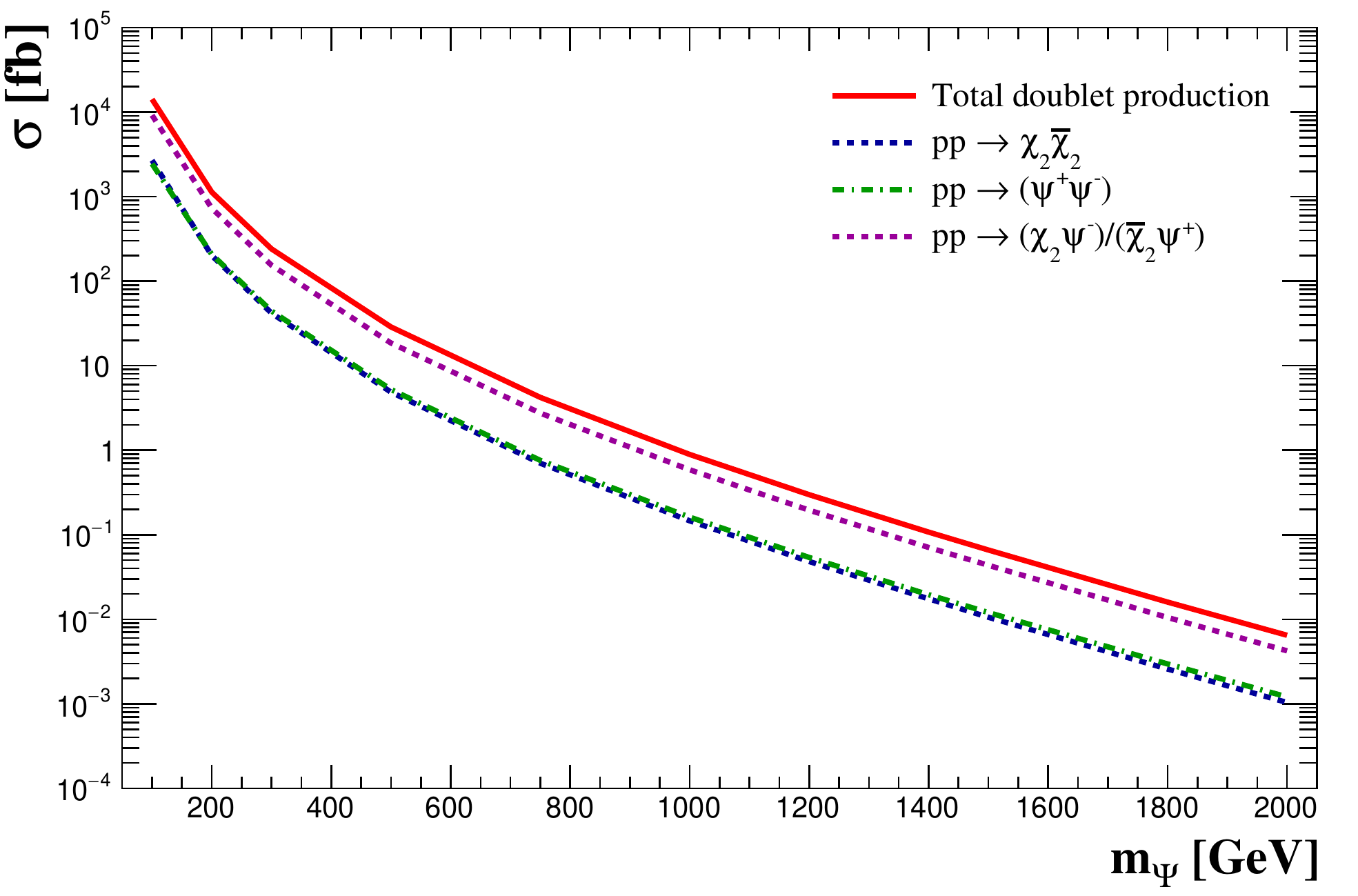}
	\includegraphics[width=0.495\linewidth]{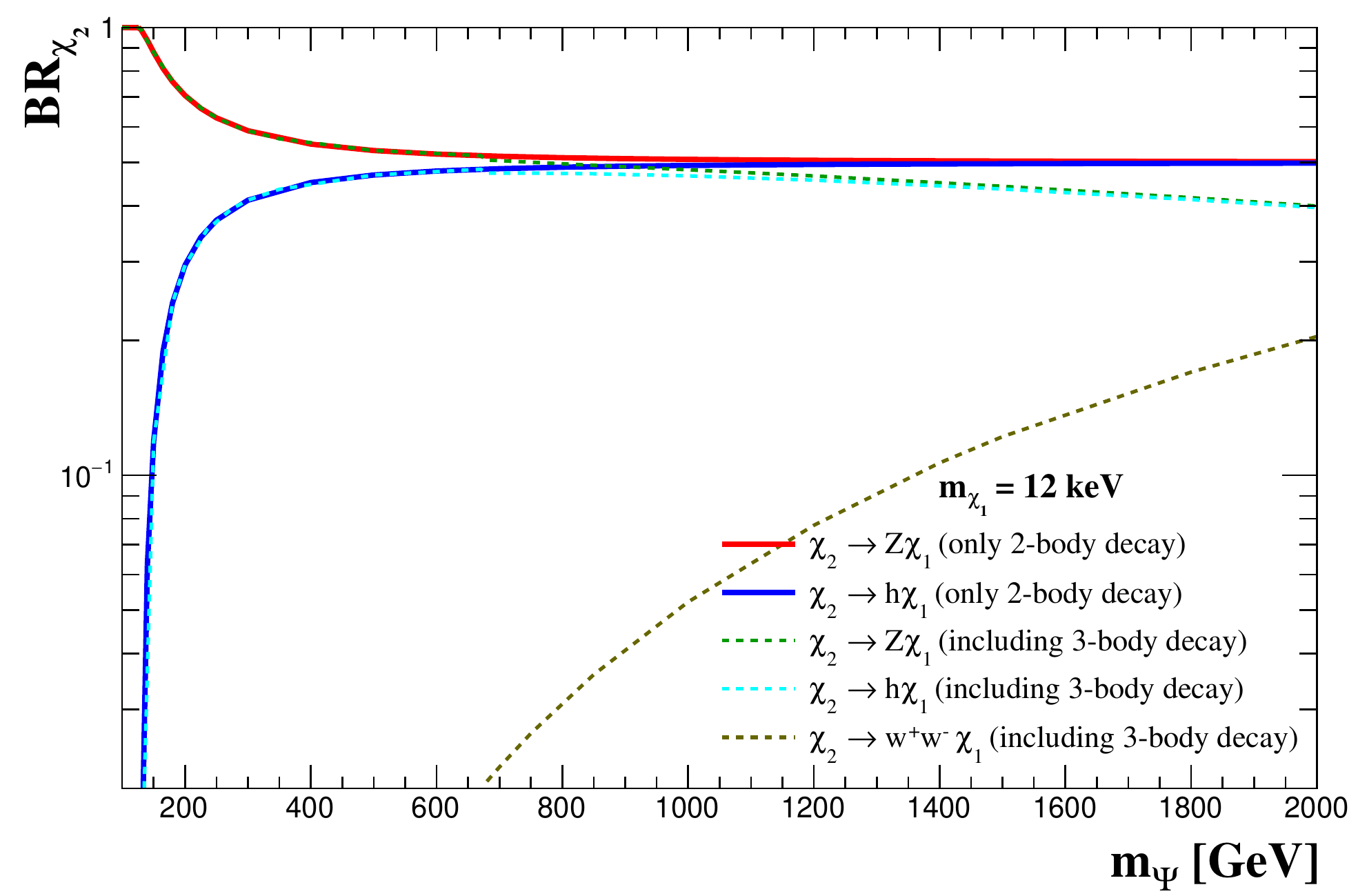}
	\caption{Left plot shows different modes of heavy BSM fermions pair production cross-section at the LHC as a function of doublet mass  $m_\psi$. In the right plot branching ratio of different decay modes of heavy neutral fermion $\chi_2$ are shown as a function of fermion doublet mass $m_\psi$.}
	\label{fig:validation_plot}
\end{figure}

\begin{figure}[!t]
	\centering
	\includegraphics[width=0.5\linewidth]{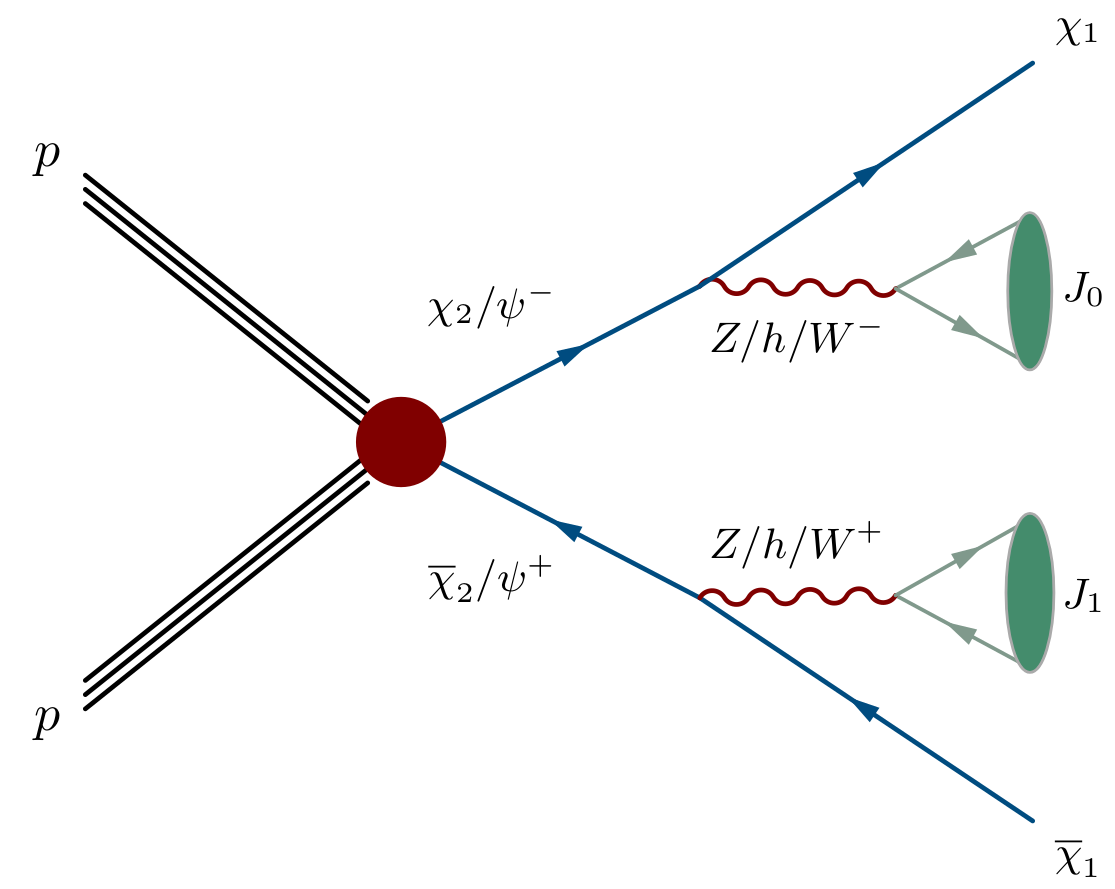}
	\caption{Representative diagram for the production of heavy BSM fermions pair at the LHC, finally generating a pair of DM candidates along with massive bosons. Hadronic decay of these bosons produces boosted jets $J_{i=0,1}$ observed at a hadron calorimeter.}
	\label{fig:feyn_diag}
\end{figure}

The production channels are s-channel pair-production of charged and neutral fermions through Z-boson or photon exchange or associated production of charged and neutral fermions through s-channel W-bosons. 
\begin{equation}
    pp\to\chi_2\overline{\chi}_2,\;\;\;\;\;\;pp\to\psi^+\psi^-, \;\;\;\;\;\;pp\to\psi^\pm\overline{\chi}_2
\end{equation}
The production cross-section is plotted in \autoref{fig:validation_plot} as a function of the doublet mass $m_\Psi$ at 14 TeV LHC. These production channels result in final state of the form 
 \begin{align}
    \chi_2 \overline{\chi}_2& \to Z(h) Z(h) \chi_1\overline{\chi}_1,\\ 
    \psi^+\psi^-&\to W^+W^-\chi_1\overline{\chi}_1,\\ 
    \psi^\pm\chi_2&\to W^\pm Z(h)\chi_1\overline{\chi}_1
 \end{align}
These heavy fermions promptly decay into a DM candidate with a boson following Eqs.~\eqref{eq:decay_width1}, \eqref{eq:decay_width2} and \eqref{eq:decay_width3}. Depending upon the mass of this heavy fermion, the boson's outcomes can be observed in the large radius hadronic clusters in the hardon calorimeter, termed fatjets. Notice the additional advantage of looking into the dominant hadronic channel if the QCD background is controlled effectively.
So, essentially our signal topology will assume the form
 \begin{equation}
     pp\to2J_{\rm{CA8}}+\slashed{E}_T
 \end{equation}
 where, $J_{\rm{CA8}}$ are jets clustered with Cambridge-Aachen algorithm with jet radius $R=0.8$, whose 2-prong substructure we intend to probe. The Feynman diagram for the signal processes is shown in \autoref{fig:feyn_diag} where $J_0$ and $J_1$ are the two fatjets originating from the hadronic decay of the gauge bosons $Z/h/W$. 
 

\subsection{Analysis setup and event simulation}
We implement the model in \texttt{FeynRules}~\cite{Alloul:2013bka} to obtain the \texttt{UFO}~\cite{Degrande:2011ua} library, which we feed to \texttt{MadGraph5v2.9.9}~\cite{Alwall:2014hca} to generate the events at leading order at $14$~TeV CM energy. \texttt{Pythia8}~\cite{Bierlich:2022pfr} is used for showering and hadronization with matching up to two to four extra jets (depending upon the processes) with MLM matching scheme~\cite{Mangano:2006rw} with virtuality-ordered parton showers. For the hadronization purpose we have used the default settings of the Pythia8. The detector effects are simulated using \texttt{Delphes3}~\cite{deFavereau:2013fsa} with the default CMS configuration card. The particle-flow objects are then clustered using \texttt{FastJet}~\cite{Cacciari:2011ma} with the anti-$k_T$ algorithm~\cite{Cacciari:2008gp}  with the radius-parameter $R=0.8$ and a minimum transverse momentum of $180$ GeV to form the fatjets. We then reclustered these jets with the Cambridge-Aachen (C/A) algorithm~\cite{Dokshitzer:1997in} keeping all the constitutents to form the fatjets to use in the subsequent analyses. We preferred here the C/A algorithm over the $k_T$ and anti-$k_T$ algorithm since it selects subjets closest to the hard jet axis compared to the other two jets~\cite{CMS:2009lxa}. It is important to discuss here the pileup and underlying events as they play an important role in the simulation of the datasets particularly in the context of HL-LHC where upto 200 interactions are expected per bunch crossing. However with the state-of-the-art techniques such as charged-hadron subtraction (CHS)~\cite{CMS:2017yfk}, pileup per particle identification (PUPPI)~\cite{Bertolini:2014bba}, SoftKiller~\cite{Cacciari:2014gra}, the effect of pileup can be substantially mitigated. In \cite{CMS:2020ebo} it was shown that for jet mass and substructure variables, missing transverse energy (MET) resolution the PUPPI algorithm gives the best performance for events with more than 30 interactions. In this work, we have assumed an ideal scenario where the effects of PU and UE can be substantially mitigated and hence did not consider these effects in the analysis.

\subsection*{Backgrounds}
Different SM processes can imitate the signature of di-fat-jets with missing transverse momentum. We extensively generated and studied these backgrounds from the following processes.
\begin{itemize}
    \item \textbf{V+jets:}  QCD jets mimic the boosted jet with the invisible decay of vector boson. Contribution from this background can be extremely large. 
    \begin{enumerate}
        \item[(i)] Z+jets -- This process gives the most dominant background where the Z-boson decays invisibly, resulting in large missing energy. 
        \item[(ii)] W+jets -- This process also contributes considerably when the W-boson decays leptonically, and the leptons are missed in the detector. These missed leptons, along with the neutrinos, contribute to missing energy. 
    \end{enumerate}
    \item \textbf{VV+jets:}  This background can be divided into three types, namely, $WZ, WW$ and $ZZ$ matched up to two extra jets. 
    \begin{enumerate}
        \item[(i)] WZ+jets -- This is the most dominant of the three where $W$-boson decays hadronically and $Z$-boson decays invisibly. 
        \item[(ii)] WW+jets -- This background contributes when the leptons from the semileptonic decay of two W-bosons are missed in the central tracker. 
        \item[(iii)] ZZ+jets -- In this process, one of the Z-bosons decays hadronically while the other invisibly. 
    \end{enumerate}
    \item \textbf{Single top:}  Three processes, namely $tW, \,tb, \, tj$, contribute to this background. Of the three single top quark production associated with a W-boson contributes most significantly. 
    \item \textbf{$t \bar{t}$+jets:}  Semileptonic decay of top-pair contributes as a significant background to our signal process. This background process can be controlled substantially by applying a $b$-tag veto. 
\end{itemize}

%
\subsection{Event Selection} \label{sim_bkg}
The detector simulated datasets are analyzed in the \texttt{ROOT}~\cite{BRUN199781} framework, where we use the following baseline-selection criteria.
\begin{itemize}
    \item Events are selected with at least two fatjets with cone-radius $R=0.8$ and transverse momentum $p_T(J_i)>180$ GeV. 
    \item The missing transverse momentum is required to be larger than $100$ GeV.
    \item Events that have lepton with transverse momentum $p_T(\ell)>10$ GeV within pseudorapidity range $|\eta(\ell)|<2.4$ are vetoed. This veto controls SM events with gauge boson decaying leptonically.
    \item We also apply a condition that the azimuthal separation between any of the fatjets and missing momentum to be less than 0.2, i.e., $|\Delta\phi(\slashed{p}_T, J_{0,1})|>0.2$. This eliminates possible contribution to missing transverse momenta coming from jet mismeasurement.
\end{itemize}
On top of the baseline-selection criteria, we applied some extra selection cuts to reduce the overwhelming backgrounds before going for the multivariate analysis (MVA) using the \texttt{TMVA} package~\cite{Hocker:2007ht} integrated into the \texttt{ROOT} framework. The additional cuts are in the following:
\begin{itemize}
    \item Events containing b-tagged small radius jets are rejected. These slim jets are clustered with an anti-$k_T$ algorithm with the radius parameter $R=0.4$. This cut particularly helps in reducing the top-pair background. 
    \item Events with the mass of both the pruned~\cite{prune} fatjets higher than $40$ GeV are accepted. Although these fatjets are expected to be even heavier, acceptance cuts are kept low, assuming further improvements in machine learning models. 
\end{itemize}

%
\subsection{Substructure Variables}
Jet substructure observables have proved to be very efficient tools for analyzing datasets in the boosted topology. The heavier particles in our model, $\psi^\pm, \chi_2$, while decaying, impart a significant boost to the daughter gauge bosons ($h, Z, W^\pm$), which can be reconstructed by clustering its hadronic decay signature with larger cone radius with a sequential algorithm. These clusters, when analyzed with suitable observables, reveal their two-prong substructure. In this analysis, we have used the following substructure variables,
\begin{figure}[t] 
    \centering
    \includegraphics[width=0.99\linewidth]{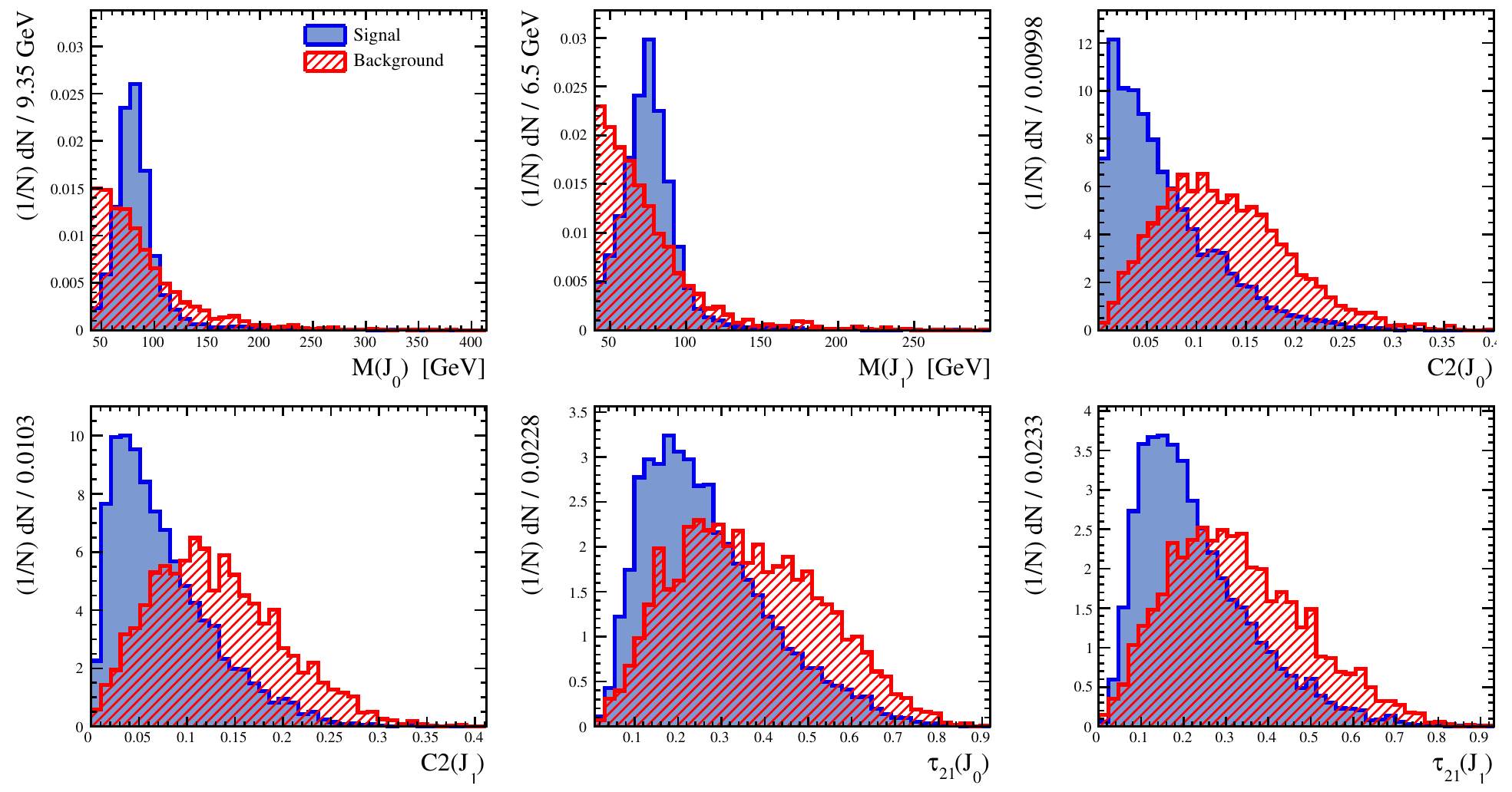}
    \includegraphics[width=0.99\linewidth]{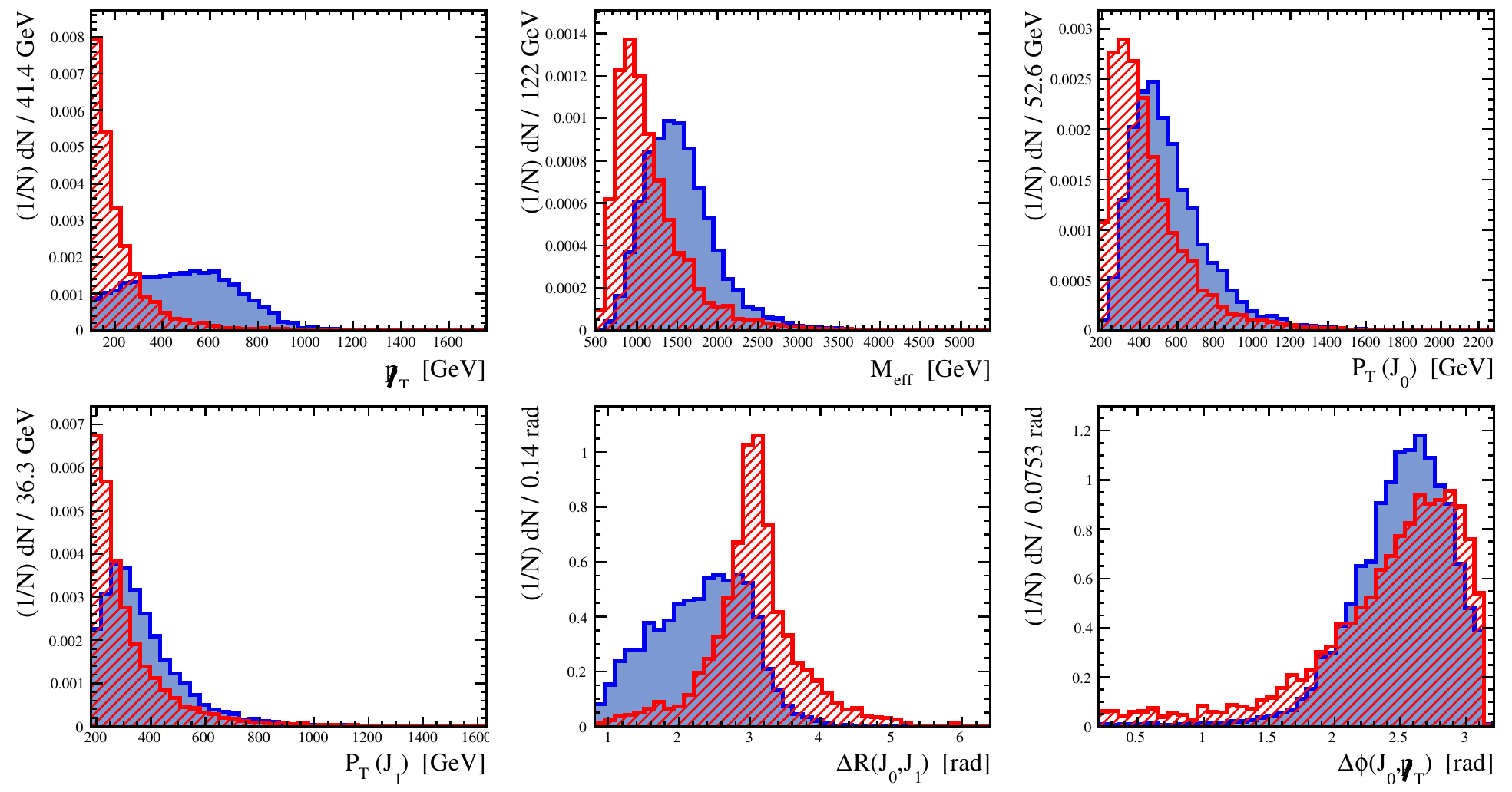}
    \caption{Normalized distributions of the important input high-level variables characterizing fat-jets used in the multivariate analysis for the discrimination of signal (blue-shaded lines) and the backgrounds (red-hatched lines) considering the first signal benchmark point BP-1.}
    \label{fig:variables}
\end{figure}
\begin{itemize}
    \item Pruned jet mass --- Systematic effects in the substructure sourced from the recombination algorithm can often be significant. Application of pruning~\cite{prune} improves identification of heavy particle decays by removing soft and wide angle proto-jets. Pruning betters the mass resolution of the heavy particles. Softdrop~\cite{Larkoski:2014wba} grooming technique also serves similar purpose, and we compared both the techniques in our analysis. In our case we found that the pruned jet mass variable performs slightly better compared to the softdropped jet mass. We performed the pruning with the standard method and as implemented in the Delphes modules with $z_{\rm{cut}}=0.1$\footnote{\texttt{RcutPrun} in \texttt{Delphes3}} and $D_{\rm{cut}}=0.5$ and checked for the condition $z={\rm min}(p_{T_i},p_{T_j})/p_{T_{i+j}}<z_{\rm cut}$ and $\Delta R_{ij}>D_{\rm{cut}}$. If the condition is satisfied, the merging is vetoed. Otherwise, a pruned jet is formed. In this analysis, we have used the pruned jet mass of the two fatjets, i.e. $M(J_0)$ and $M(J_1)$, whose distribution is shown in \autoref{fig:variables}.
    \item Energy correlator --- The generalized energy correlation functions have proved to be very efficient in probing the N-prong substructure of a boosted jet just using the energy and pairwise angles of the particles without requiring explicit identification of sub-jet regions. We use here the \emph{energy correlation double ratio} defined as~\cite{Larkoski:2013eya}, 
    \begin{equation}
        C_N^{(\beta)}=\frac{{\rm ECF}(N+1,\beta){\rm ECF}(N-1,\beta)}{{\rm ECF}(N,\beta)^2}
    \end{equation}
    where, ${\rm ECF}(N,\beta)$ is defined as 
    \begin{equation*}
        {\rm ECF}(N,\beta)=\sum_{i_1<i_2<...<i_N\in J}\left(\prod_{a=1}^N p_{T_{i_a}}\right) \left(\prod_{b=1}^{N-1}\prod_{c=b+1}^N R_{i_bi_c}\right)
    \end{equation*}
    $R_{ij}$ being the distance between $i$ and $j$ in the rapidity-azimuthal angle plane, $R_{ij}=\sqrt{(y_i-y_j)^2+(\phi_i-\phi_j)^2}$ and $\beta$, the angular exponent that helps in optimization of the discrimination power. As discussed in \cite{Larkoski:2013eya}, for 2-prong substructures like the $W/Z/h$, as in our case, one needs a 3-point energy correlator, i.e., $C_2^\beta$. So we used the variables $C_2(J_0)$ and $C_2(J_1)$ respectively for the leading and sub-leading fatjets.

    \item N-subjettiness ratio --- N-subjettiness is an excellent jet-shape variable for boosted object identification. This variable is essentially a measure of how well the jet energies are aligned into different constituent subjets in an N-prong fatjet. It is defined as~\cite{Thaler:2010tr},
\begin{equation}
    \tau_N^{(\beta)}=\frac{1}{N_0}\sum_k p_{T_k}\, \rm{min(\Delta R_{1,k},\Delta R_{2,k},...,\Delta R_{N,k})}.
\end{equation}
    Here, $k$ runs over constituent particles in a given jet and $\Delta R_{J,k}=\sqrt{(\Delta\eta)^2+(\Delta\phi)^2}$ is the distance in the $\eta-\phi$ plane between a candidate subjet J and a constituent particle k. The normalization factor is defined as $N_0=\sum_k p_{T_k} R_0$, where $p_{T_k}$ is the transverse momentum of the jet of radius $R_0$. As shown in \cite{Thaler:2010tr}, rather than $\tau_N$, a better variable can be the ratio $\tau_{N}/\tau_{N-1}$ for identification of an N-prong fatjet, so we chose the subjettiness ratio $\tau_{21}=\tau_2/\tau_1$ of the two fatjets as part of our substructure variables in the analysis.
\end{itemize}

Other than these substructure variables, we use other kinematic variables that appear sensitive in isolating the signal from the vast background. 
\begin{itemize}
    \item The missing transverse momentum, $\slashed{p}_T$.
    \item The effective mass of the process $M_{eff}=\sum_{vis}|p_T|+|\slashed{p}_T|$.
    \item Transverse momentum of the leading and sub-leading fatjets, $p_T(J_0)$, $p_T(J_1)$ %
    \item The distance between the two fatjets in the $\eta-\phi$ plane, $\Delta R(J_0,J_1)$.
    \item The difference in the azimuthal angle between the missing transverse momentum and the leading fatjets, $\Delta\phi(\slashed{p}_T, J_0)$.
\end{itemize}
In \autoref{fig:variables}, the normalized distributions of all the variables are shown where the first two rows contain the substructure variables, and the other two rows present the rest of the variables. The distributions for the background (red hatched) consists of weighted contributions from all the variables discussed in \autoref{sim_bkg}. 
\begin{figure}[t] 
	\centering
	\includegraphics[width=0.55\linewidth]{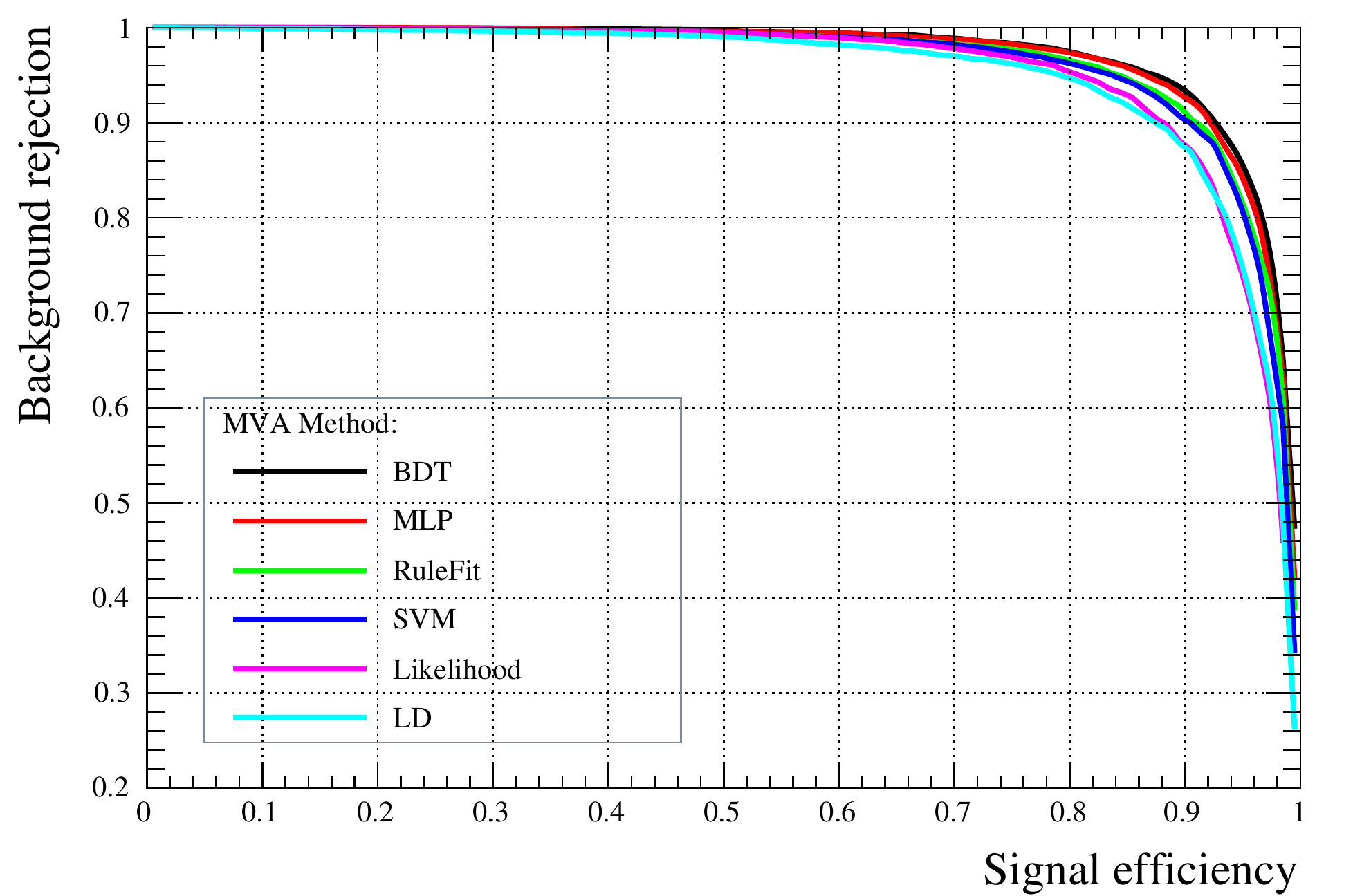}
	\caption{Receiver Operating Characteristic curves showing all the MVA methods considered for optimization for benchmark point BP-1.}
	\label{fig:ROC}
\end{figure}

\begin{table}[t] 
	\small
	\centering
	\begin{tabular}{|l|c|}
		\hline
		\texttt{BoostType} & AdaBoost \\
		\texttt{AdaBoostBeta} & 0.34 \\
		\texttt{NTrees} & 390  \\
		\texttt{MaxDepth} & 4 \\
		\texttt{MinNodeSize} & 5\% \\
		\texttt{BaggedSampleFraction} & 0.6 \\
		\texttt{nCuts} & 10 \\
		\texttt{SeparationType} & GiniIndex \\
		\texttt{PruneMehod} & CostComplexity \\
		\texttt{PruneStrenth} & 0.5 \\
		\hline
	\end{tabular}
	\caption{List of the parameters used in the adaptive BDT analysis.}
	\label{tab:bdt_param}
\end{table}
\subsection{Multivariate Analysis}
We analyze the datasets using the multivariate analysis (MVA) method, which produces non-linear decision boundaries more efficiently than the univariate cut-based analysis. We employ the adaptive boosted decision tree (BDT) method among the other different MVA methods since it results in better performance in terms of the Receiver Operating Characteristic (ROC). In \autoref{fig:ROC}, we have shown the ROC curves of all the methods that we have considered while optimizing the training of the signal and background datasets for the MVA analysis. The configuration we used is shown in \autoref{tab:bdt_param}. 

\begin{figure}[t] 
	\centering
	\includegraphics[width=0.49\linewidth]{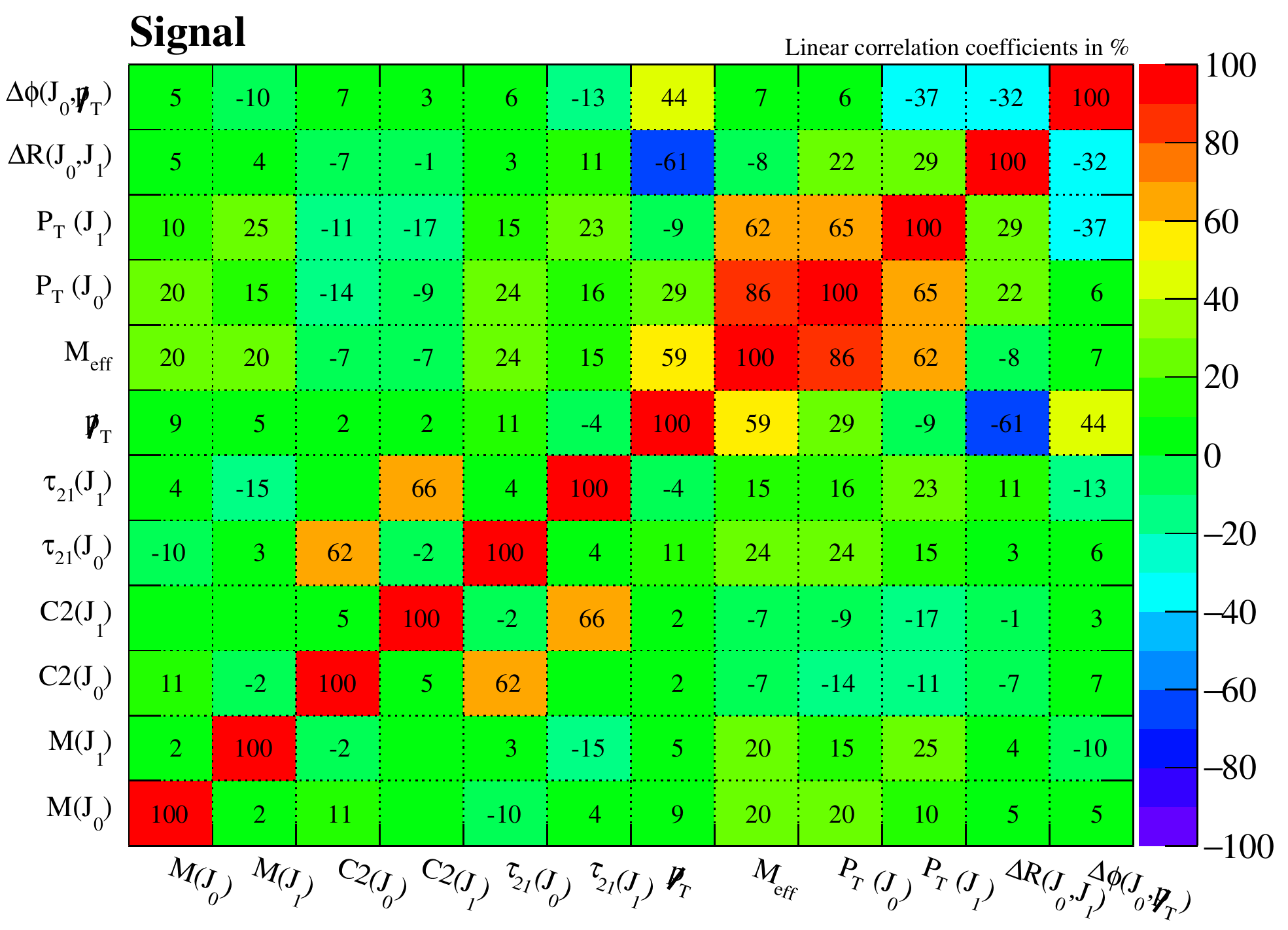}
	\includegraphics[width=0.49\linewidth]{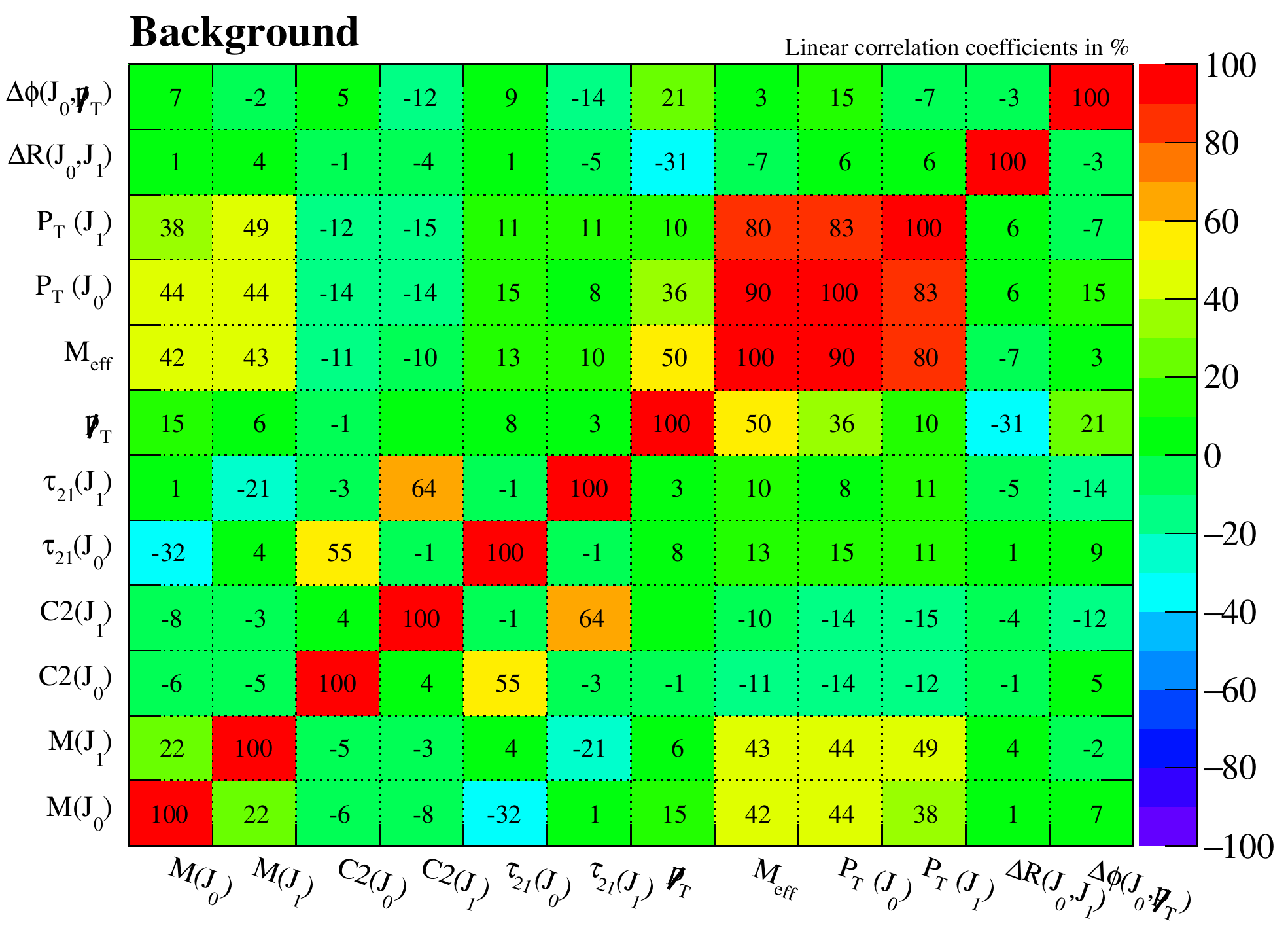}
	\caption{Plots showing the Linear Correlation between the input variables used in the MVA analysis for signal (left panel) and background (right panel) for benchmark point BP-1.}
	\label{fig:correlation}
\end{figure}

The preprocessing of the datasets is done by selecting a subset of variables from a larger collection of variables. Such choice is based on the linear correlation among them and their relative importance as a discriminator between the signal and background. For evaluating the linear correlation coefficient, we used the formula 
\begin{equation}
	\rho(x,y)=\frac{\rm{cov}(x,y)}{\sigma_x\sigma_y}
\end{equation}
where, $\rm{cov}(x,y)=\langle xy \rangle - \langle x\rangle\langle y\rangle $ is the covariance between $x$ and $y$ and $\sigma_x$ ($\sigma_y$) is the standard deviation of the the variable $x$ ($y$). We illustrated the correlation between different variables in \autoref{fig:correlation} for signal (left panel) and background (right panel) where we see that most of the variables used are uncorrelated. Some variables like $p_T(J_0)$, $M_{eff}$ show a higher correlation between them.  But considering their high separation power, we have still used them. In \autoref{tab:var_rank}, the relative importance of the variables for discriminating the signal from the background in the classification process. 

\begin{table}[t] 
	\centering
	\resizebox{\textwidth}{!}{%
		\begin{tabular}{|c|c|c|c|c|c|c|c|c|c|c|c|}
			\hline
			\Gape[6pt][3pt]{$\slashed{p}_T$} & $\Delta R(J_0,J_1)$ & $C_2(J_0)$ & $M_{eff}$ & $C_2(J_1)$ & $M(J_0)$ & $M(J_1)$ & $p_T(J_0)$ & $\tau_{21}(J_1)$ & $\tau_{21}(J_0)$ & $p_T(J_1)$ & $\Delta\phi(J_0,\slashed{p}_T)$ \\ \hline
			\Gape[6pt][5pt]{44.41} & 26.90 & 24.29 & 22.79 & 19.20 & 14.94 & 13.23 & 12.23 & 11.02 & 9.30 & 9.20 & 4.50 \\ 
			\hline
		\end{tabular}
	}
	\caption{Method unspecific relative separation power of the different input variables according to their rank before using at MVA.}
	\label{tab:var_rank}
\end{table}

\begin{figure}[t] 
    \centering
    \includegraphics[width=0.49\linewidth]{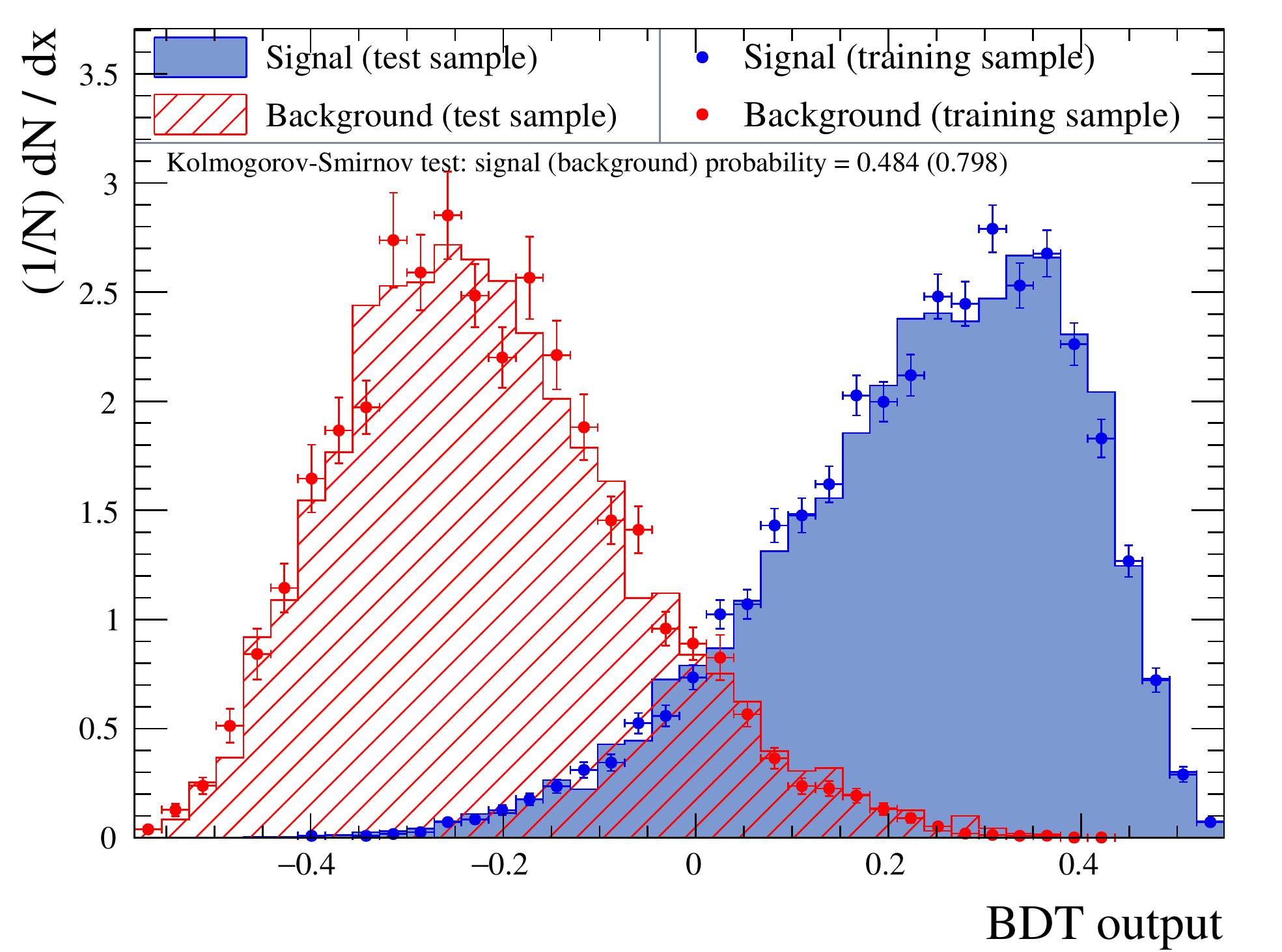}
    \includegraphics[width=0.49\linewidth]{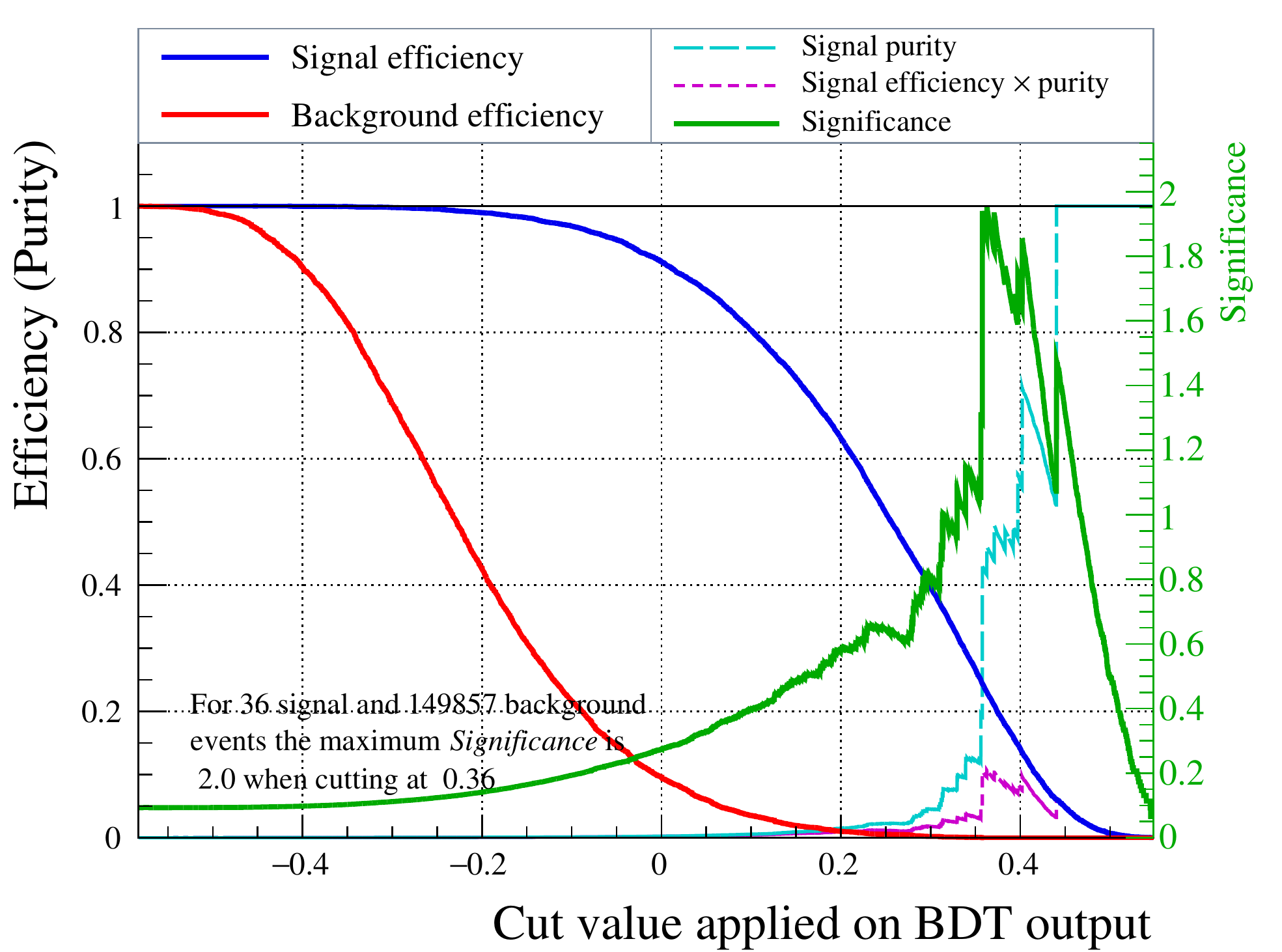}
    \caption{Normalized distributions of the BDT output (left panel) for the signal in the background with the result of the overtraining test (KS value) and efficiency curve (right panel) for signal and background as a function of the BDT cut value. We also show the variations of the significance obtained in the analysis with the cut value. Both plots are generated considering the first benchmark point BP-1.}
    \label{fig:overtrain&sig}
\end{figure}

The parameter setup we used for BDT training is given in \autoref{tab:bdt_param}. Despite being a robust classifier, BDT can suffer from overtraining. We have checked this by looking at the Kolmogorov-Smirnov probability as shown in the left panel of \autoref{fig:overtrain&sig} along with BDT output for the benchmark point. In the right panel of \autoref{fig:overtrain&sig}, we have shown the signal (background) efficiency (i.e. the fraction of the selected events from the total events fed to the classifier) in blue (red) line, signal purity (evaluated as $S/(S+B)$, with $S$ and $B$ being the signal and background events after BDT cut) in cyan line and the signal significance in green line as functions of BDT cut values. We have evaluated the signal significance using the formula, $sig=S/\sqrt{S+B}$, where $S$ and $B$ are the signal and background events after the application of the BDT cut. We see that a significance of 2.0 is achieved by applying a cut at $\sim0.36$ for the benchmark points considered. Note that the value of the optimal cut may differ for a different point in the parameter space we are working.

%
\subsection{Results}\label{results}
In this section, we have presented the results of our MVA analysis. As discussed in \autoref{model}, the main independent parameters of the model are the doublet mass ($m_\psi$), coupling ($y$) and the dark matter mass ($m_\chi$). So for a fixed dark matter mass, the parameter space of the model can be defined with $y$ and $m_\Psi$. Based on these parameters, we considered two non-standard benchmark points as shown in \autoref{tab:BP}
to demonstrate the performance of our BDT analysis in two different regions based on accessibility by previous searches. One of those points lies in a region probed by the displaced vertex search performed in ref.~\cite{Calibbi:2018fqf}, whereas the other point lies outside the region excluded by the displaced vertex search. 

\begin{table}[t] 
\centering
\resizebox{\textwidth}{!}{%
\begin{tabular}{|c|c|c|c|c|c|c|c|c|}
\hline
\Gape[5pt][5pt]{Type} & Z+jets & W+jets & ZZ+jets & WZ+jets & WW+jets & single-top & $t\overline{t}+jets$ & Total \\
\hline
Events & \multirow{2}{5em}{\centering 71136} & \multirow{2}{5em}{\centering 52952} & \multirow{2}{5em}{\centering 4226} & \multirow{2}{5em}{\centering 2101} & \multirow{2}{5em}{\centering 593} & \multirow{2}{5em}{\centering 4452} & \multirow{2}{5em}{\centering 14398} & \multirow{2}{5em}{\centering 149857} \\
before BDT & & & & & & & & \\
\hline\hline
\multirow{2}{5em}{\centering Benchmark} & Events & Optimal & Background & Background & Signal & Signal & \multicolumn{2}{c|}{\multirow{2}{5em}{\centering Significance}} \\
& before BDT & BDT cut & efficiency & events & efficiency & events & \multicolumn{2}{c|}{}\\
\hline
\Gape[5pt][5pt]{BP-1} & 36 & 0.363 & 6.46$\times10^{-5}$ & 10 & 0.2306 & 8 & \multicolumn{2}{c|}{2.0} \\
\hline
\Gape[5pt][5pt]{BP-2} & 45 & 0.359 & 8.36$\times10^{-5}$ & 13 & 0.2129 & 10 & \multicolumn{2}{c|}{2.0} \\
\hline
\end{tabular}
}
\caption{Cut-flow table for the signal and backgrounds. The first two rows give the event numbers before applying the BDT cut, and the rest of the rows contain information on signal events before and after the BDT cut (along with the cut efficiencies), the optimal BDT cut values, the background events after BDT cut along with the cut efficiency and the obtained significance for the two benchmark points at $\mathcal{L}_{int}=300$fb$^{-1}$.}
\label{tab:result}
\end{table}
In the first two rows of \autoref{tab:result}, we gave the event yields for each background before BDT analysis and then we presented the event yields for signal (both benchmark points) along with the numbers after BDT analysis, the optimal BDT cut value and the obtained significance.  

\begin{figure}[t]
	\centering
	\includegraphics[width=0.9\textwidth]{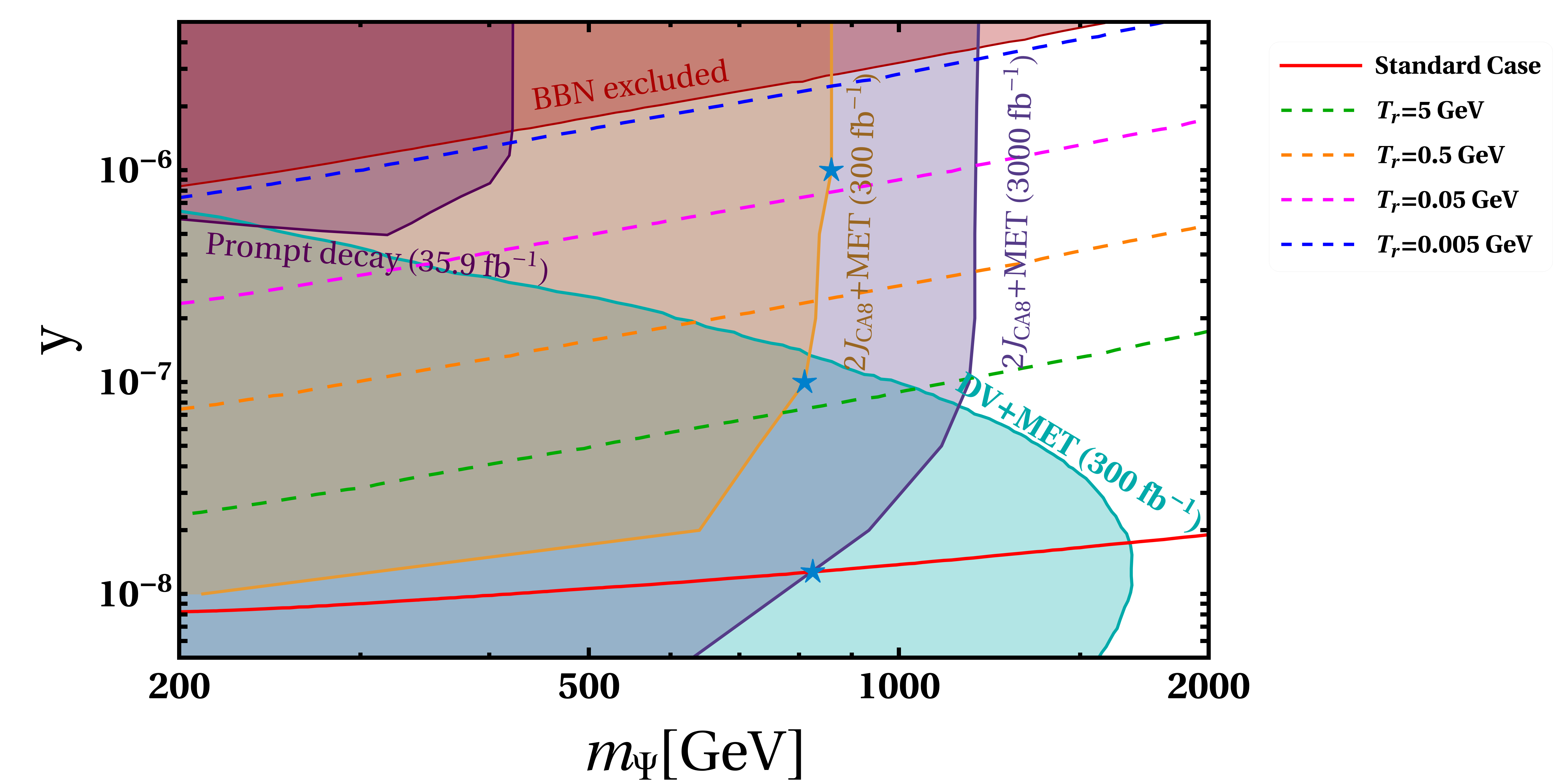}
	\caption{Exclusion limits at 14 TeV c.m. energy at LHC with $\mathcal{L}_{int}=300$ fb$^{-1}$ (orange-shaded region) and $3000$ fb$^{-1}$ (blue-shaded region) for a DM mass of 12 keV. The shaded region in cyan and purple are excluded by the previous displaced vertex search (with $300$ fb$^{-1}$) and prompt search at LHC (with 35.9 fb$^{-1}$). Also, contour lines from \autoref{fig:DMrelic} are also shown to indicate the permissible region from the relic density constraint in the non-standard cosmological scenario in the singlet doublet dark matter model parameter space.	The blue-colored stars represents the three benchmark points discussed in \autoref{DMpheno}.}
	\label{fig:exclusion_plt}
\end{figure}

Based on this optimized setup we scanned the model parameter space to obtain a $2\sigma$ exclusion limit. We note that the production of the heavy mediators ($\chi_2,\;\psi^\pm$) occurs through electroweak processes, but the dependence of the coupling $y$ in the production cross-section of the dark matter candidate ($\chi_1$) is induced via the decay widths (or, branching fractions) of the mediators as shown in Eqs.~\eqref{eq:decay_width1},\eqref{eq:decay_width2} and \eqref{eq:decay_width3}. 
In \autoref{DMpheno}, we discussed that the dark matter phenomenology can be presented in the plane of the coupling ($y$) and the doublet mass ($m_\psi$). In \autoref{fig:exclusion_plt}, we compiled our results from the collider analysis along with the results from DM phenomenology in the same plane to estimate the current limits on our model parameters in the prompt search context. In \autoref{fig:exclusion_plt}, the region on the left of the brown dashed contour represents the region excluded by our analysis at $300$ fb$^{-1}$ of luminosity. We also presented the results from the prompt search as performed in \cite{CMS:2018szt} in the context of supersymmetric charginos and neutralinos in the purple dashed contour. For comparison, we also presented the analysis results from the displaced vertex + missing momentum search from \cite{Calibbi:2018fqf}. We see that our analysis can probe a larger area in the parameter space when the coupling is substantially large to trigger a prompt decay of the mediators. We also presented the result for a luminosity of $3000$ fb$^{-1}$ to project the searching prospect of HL-LHC. We found that while at $300$ fb$^{-1}$ luminosity the bound on the doublet mass can reach up to $880$ GeV, with the increased luminosity of 3000 fb$^{-1}$ it goes beyond $1235$ GeV.

%
\section{Summary and conclusion}\label{conclusion}
In this work, we examine a simple extension of the SM with a singlet and a doublet fermion, where the lightest neutral particle serves as a dark matter candidate. Due to its gauge interaction, the doublet part thermalizes with the SM particles in the early Universe. On the contrary, the dark matter candidate cannot do the same and remains in the primordial soup with negligible or no presence. In this scenario, gradual dark matter production occurs via the decay of heavy doublet fermion through a tiny Yukawa coupling or the singlet doublet mixing angle such that the DM never reaches thermal equilibrium. Non-thermal freeze-in dark matter is one of the exciting possibilities explaining the non-observation of any signature yet at the different ground or space-based DM detection experiments, owing to this feeble interaction.

Interestingly, the long-lived particle with displaced vertex signature provides some handle for the BSM models with the heavy mediator and light dark matter. In addition, the lightweight dark matter with typical mass scale $\mathcal{O}(10)$ keV can alleviate the small scale issue of $\Lambda$CDM by suppressing the structure formation at that scale as a result of the warmness of dark matter. Moreover, such dark matter offers the platform to be tested in various cosmological and astrophysical probes.

Lacking evidence of the energy content of the Universe in the pre-BBN era, we consider the possibility of non-standard cosmology and discuss its effect on the evolution of dark matter. We have shown that the fast-expanding Universe demands a larger interaction rate than the standard scenario (radiation-dominated Universe) to satisfy the relic density constraint. 
 
As a fallout of this requirement, not only the dark matter phenomenology but one needs to relook at the corresponding collider search strategy. Characteristic long-lived particle searches at the collider experiment and the displaced vertex signature can no more constrain this scenario effectively. In this work, we construct the extended parameter space that can satisfy the dark matter relic density and build a suitable search at the LHC that can effectively explore a large part of these parameters. 

After copious production of dark sector heavy fermions at the LHC, they promptly decay into the dark matter candidate associated with a heavy gauge boson or a Higgs boson. The heaviness of these mother particles with lightweight dark matter ensures the production of boosted hadronic jets from the final decay of such bosons. We capitalize on such inherent features to build our search strategy comprised of two boosted jets and the substantial MET. These boosted fat jets bears characteristic substructure pattern different from single prong large radius QCD jets, which can help us probe through this channel, prevailing over the extensive QCD background.
We have also adopted multivariate analysis based on the boosted decision tree with the help of twelve kinematic and boosted jet variables to provide the optimum separation between signal and background. 
Our analysis puts severe constraints on the vast plane of Yukawa coupling and heavy particle mass at 14 TeV LHC, for $\mathcal{L}_\text{int}$ = 300 fb$^{-1}$ and 3000 fb$^{-1}$. Most of the region of our parameter space remains inaccessible for the displaced vertex constraint.

\acknowledgments
 This work is supported by the Physical Research Laboratory (PRL), Department of Space, Government of India. Part of the computational work is performed using the HPC resources (Vikram-100 HPC) and TDP project at PRL. Part of the work of SK is performed using the HPC resources (Koshambi) at the BITS Pilani, K.K. Birla Goa Campus, mentained by the CSIS department. SK would like to acknowledge the fellowship support of the BITS Pilani.

\bibliographystyle{JHEP}
\bibliography{Reference.bib}

\providecommand{\href}[2]{#2}\begingroup\raggedright\begin{thebibliography}{100}

\bibitem{Sofue:2000jx}
Y.~Sofue and V.~Rubin, \emph{{Rotation curves of spiral galaxies}},
  \href{http://dx.doi.org/10.1146/annurev.astro.39.1.137}{\emph{Ann. Rev.
  Astron. Astrophys.} {\bfseries 39} (2001) 137--174},
  [\href{https://arxiv.org/abs/astro-ph/0010594}{{\ttfamily
  astro-ph/0010594}}].

\bibitem{Clowe:2006eq}
D.~Clowe, M.~Bradac, A.~H. Gonzalez, M.~Markevitch, S.~W. Randall, C.~Jones
  et~al., \emph{{A direct empirical proof of the existence of dark matter}},
  \href{http://dx.doi.org/10.1086/508162}{\emph{Astrophys. J. Lett.} {\bfseries
  648} (2006) L109--L113},
  [\href{https://arxiv.org/abs/astro-ph/0608407}{{\ttfamily
  astro-ph/0608407}}].

\bibitem{Hinshaw_2013}
G.~Hinshaw, D.~Larson, E.~Komatsu, D.~N. Spergel, C.~L. Bennett, J.~Dunkley
  et~al., \emph{Nine-year wilkinson microwave anisotropy probe ( wmap )
  observations: Cosmological parameter results},
  \href{http://dx.doi.org/10.1088/0067-0049/208/2/19}{\emph{The Astrophysical
  Journal Supplement Series} {\bfseries 208} (Sep, 2013) 19}.

\bibitem{Planck:2018vyg}
{\scshape Planck} collaboration, N.~Aghanim et~al., \emph{{Planck 2018 results.
  VI. Cosmological parameters}},
  \href{http://dx.doi.org/10.1051/0004-6361/201833910}{\emph{Astron.
  Astrophys.} {\bfseries 641} (2020) A6},
  [\href{https://arxiv.org/abs/1807.06209}{{\ttfamily 1807.06209}}].

\bibitem{Green:2002ht}
A.~M. Green, \emph{{Effect of halo modeling on WIMP exclusion limits}},
  \href{http://dx.doi.org/10.1103/PhysRevD.66.083003}{\emph{Phys. Rev. D}
  {\bfseries 66} (2002) 083003},
  [\href{https://arxiv.org/abs/astro-ph/0207366}{{\ttfamily
  astro-ph/0207366}}].

\bibitem{Chang:2017gla}
C.-F. Chang, X.-G. He and J.~Tandean, \emph{{Two-Higgs-Doublet-Portal
  Dark-Matter Models in Light of Direct Search and LHC Data}},
  \href{http://dx.doi.org/10.1007/JHEP04(2017)107}{\emph{JHEP} {\bfseries 04}
  (2017) 107}, [\href{https://arxiv.org/abs/1702.02924}{{\ttfamily
  1702.02924}}].

\bibitem{Chang:2017dvm}
C.-F. Chang, X.-G. He and J.~Tandean, \emph{{Exploring Spin-3/2 Dark Matter
  with Effective Higgs Couplings}},
  \href{http://dx.doi.org/10.1103/PhysRevD.96.075026}{\emph{Phys. Rev. D}
  {\bfseries 96} (2017) 075026},
  [\href{https://arxiv.org/abs/1704.01904}{{\ttfamily 1704.01904}}].

\bibitem{Visinelli:2017qga}
L.~Visinelli, \emph{{(Non-)thermal production of WIMPs during kination}},
  \href{http://dx.doi.org/10.3390/sym10110546}{\emph{Symmetry} {\bfseries 10}
  (2018) 546}, [\href{https://arxiv.org/abs/1710.11006}{{\ttfamily
  1710.11006}}].

\bibitem{Arcadi:2017wqi}
G.~Arcadi, M.~Lindner, F.~S. Queiroz, W.~Rodejohann and S.~Vogl,
  \emph{{Pseudoscalar Mediators: A WIMP model at the Neutrino Floor}},
  \href{http://dx.doi.org/10.1088/1475-7516/2018/03/042}{\emph{JCAP} {\bfseries
  03} (2018) 042}, [\href{https://arxiv.org/abs/1711.02110}{{\ttfamily
  1711.02110}}].

\bibitem{Choubey:2017yyn}
S.~Choubey, S.~Khan, M.~Mitra and S.~Mondal, \emph{{Singlet-Triplet Fermionic
  Dark Matter and LHC Phenomenology}},
  \href{http://dx.doi.org/10.1140/epjc/s10052-018-5785-1}{\emph{Eur. Phys. J.
  C} {\bfseries 78} (2018) 302},
  [\href{https://arxiv.org/abs/1711.08888}{{\ttfamily 1711.08888}}].

\bibitem{Reinert:2017aga}
A.~Reinert and M.~W. Winkler, \emph{{A Precision Search for WIMPs with Charged
  Cosmic Rays}},
  \href{http://dx.doi.org/10.1088/1475-7516/2018/01/055}{\emph{JCAP} {\bfseries
  01} (2018) 055}, [\href{https://arxiv.org/abs/1712.00002}{{\ttfamily
  1712.00002}}].

\bibitem{Evans:2017kti}
J.~A. Evans, S.~Gori and J.~Shelton, \emph{{Looking for the WIMP Next Door}},
  \href{http://dx.doi.org/10.1007/JHEP02(2018)100}{\emph{JHEP} {\bfseries 02}
  (2018) 100}, [\href{https://arxiv.org/abs/1712.03974}{{\ttfamily
  1712.03974}}].

\bibitem{Garny:2018icg}
M.~Garny, J.~Heisig, M.~Hufnagel and B.~L\"ulf, \emph{{Top-philic dark matter
  within and beyond the WIMP paradigm}},
  \href{http://dx.doi.org/10.1103/PhysRevD.97.075002}{\emph{Phys. Rev. D}
  {\bfseries 97} (2018) 075002},
  [\href{https://arxiv.org/abs/1802.00814}{{\ttfamily 1802.00814}}].

\bibitem{Blanco:2019hah}
C.~Blanco, M.~Escudero, D.~Hooper and S.~J. Witte, \emph{{Z' mediated WIMPs:
  dead, dying, or soon to be detected?}},
  \href{http://dx.doi.org/10.1088/1475-7516/2019/11/024}{\emph{JCAP} {\bfseries
  11} (2019) 024}, [\href{https://arxiv.org/abs/1907.05893}{{\ttfamily
  1907.05893}}].

\bibitem{Bhardwaj:2018lma}
A.~Bhardwaj, A.~Das, P.~Konar and A.~Thalapillil, \emph{{Looking for Minimal
  Inverse Seesaw scenarios at the LHC with Jet Substructure Techniques}},
  \href{http://dx.doi.org/10.1088/1361-6471/ab7769}{\emph{J. Phys. G}
  {\bfseries 47} (2020) 075002},
  [\href{https://arxiv.org/abs/1801.00797}{{\ttfamily 1801.00797}}].

\bibitem{Bhardwaj:2019mts}
A.~Bhardwaj, P.~Konar, T.~Mandal and S.~Sadhukhan, \emph{{Probing the inert
  doublet model using jet substructure with a multivariate analysis}},
  \href{http://dx.doi.org/10.1103/PhysRevD.100.055040}{\emph{Phys. Rev. D}
  {\bfseries 100} (2019) 055040},
  [\href{https://arxiv.org/abs/1905.04195}{{\ttfamily 1905.04195}}].

\bibitem{Konar:2020wvl}
P.~Konar, A.~Mukherjee, A.~K. Saha and S.~Show, \emph{{Linking pseudo-Dirac
  dark matter to radiative neutrino masses in a singlet-doublet scenario}},
  \href{http://dx.doi.org/10.1103/PhysRevD.102.015024}{\emph{Phys. Rev. D}
  {\bfseries 102} (2020) 015024},
  [\href{https://arxiv.org/abs/2001.11325}{{\ttfamily 2001.11325}}].

\bibitem{Konar:2020vuu}
P.~Konar, A.~Mukherjee, A.~K. Saha and S.~Show, \emph{{A dark clue to seesaw
  and leptogenesis in a pseudo-Dirac singlet doublet scenario with
  (non)standard cosmology}},
  \href{http://dx.doi.org/10.1007/JHEP03(2021)044}{\emph{JHEP} {\bfseries 03}
  (2021) 044}, [\href{https://arxiv.org/abs/2007.15608}{{\ttfamily
  2007.15608}}].

\bibitem{Heurtier:2019beu}
L.~Heurtier and H.~Partouche, \emph{{Spontaneous Freeze Out of Dark Matter From
  an Early Thermal Phase Transition}},
  \href{http://dx.doi.org/10.1103/PhysRevD.101.043527}{\emph{Phys. Rev. D}
  {\bfseries 101} (2020) 043527},
  [\href{https://arxiv.org/abs/1912.02828}{{\ttfamily 1912.02828}}].

\bibitem{Habermehl:2020njb}
M.~Habermehl, M.~Berggren and J.~List, \emph{{WIMP Dark Matter at the
  International Linear Collider}},
  \href{http://dx.doi.org/10.1103/PhysRevD.101.075053}{\emph{Phys. Rev. D}
  {\bfseries 101} (2020) 075053},
  [\href{https://arxiv.org/abs/2001.03011}{{\ttfamily 2001.03011}}].

\bibitem{Xing:2021pkb}
C.-Y. Xing and S.-H. Zhu, \emph{{Dark Matter Freeze-Out via Catalyzed
  Annihilation}},
  \href{http://dx.doi.org/10.1103/PhysRevLett.127.061101}{\emph{Phys. Rev.
  Lett.} {\bfseries 127} (2021) 061101},
  [\href{https://arxiv.org/abs/2102.02447}{{\ttfamily 2102.02447}}].

\bibitem{Borah:2022byb}
D.~Borah, S.~Jyoti~Das, A.~K. Saha and R.~Samanta, \emph{{Probing WIMP dark
  matter via gravitational waves\textquoteright{} spectral shapes}},
  \href{http://dx.doi.org/10.1103/PhysRevD.106.L011701}{\emph{Phys. Rev. D}
  {\bfseries 106} (2022) L011701},
  [\href{https://arxiv.org/abs/2202.10474}{{\ttfamily 2202.10474}}].

\bibitem{Belanger:2022qxt}
G.~Belanger, A.~Mjallal and A.~Pukhov, \emph{{WIMP and FIMP dark matter in the
  inert doublet plus singlet model}},
  \href{http://dx.doi.org/10.1103/PhysRevD.106.095019}{\emph{Phys. Rev. D}
  {\bfseries 106} (2022) 095019},
  [\href{https://arxiv.org/abs/2205.04101}{{\ttfamily 2205.04101}}].

\bibitem{Gines:2022qzy}
E.~U. Gin\'es, O.~Mena and S.~J. Witte, \emph{{Revisiting constraints on WIMPs
  around primordial black holes}},
  \href{http://dx.doi.org/10.1103/PhysRevD.106.063538}{\emph{Phys. Rev. D}
  {\bfseries 106} (2022) 063538},
  [\href{https://arxiv.org/abs/2207.09481}{{\ttfamily 2207.09481}}].

\bibitem{Bernal:2022wck}
N.~Bernal and Y.~Xu, \emph{{WIMPs during reheating}},
  \href{http://dx.doi.org/10.1088/1475-7516/2022/12/017}{\emph{JCAP} {\bfseries
  12} (2022) 017}, [\href{https://arxiv.org/abs/2209.07546}{{\ttfamily
  2209.07546}}].

\bibitem{Kundu:2021cmo}
S.~Kundu, A.~Guha, P.~K. Das and P.~S.~B. Dev, \emph{{EFT analysis of
  leptophilic dark matter at future electron-positron colliders in the
  mono-photon and mono-Z channels}},
  \href{http://dx.doi.org/10.1103/PhysRevD.107.015003}{\emph{Phys. Rev. D}
  {\bfseries 107} (2023) 015003},
  [\href{https://arxiv.org/abs/2110.06903}{{\ttfamily 2110.06903}}].

\bibitem{Medina:2021ram}
A.~D. Medina, N.~I. Mileo, A.~Szynkman and S.~A. Tanco, \emph{{Elusive muonic
  WIMP}}, \href{http://dx.doi.org/10.1103/PhysRevD.106.075018}{\emph{Phys. Rev.
  D} {\bfseries 106} (2022) 075018},
  [\href{https://arxiv.org/abs/2112.09103}{{\ttfamily 2112.09103}}].

\bibitem{Tallman:2022nts}
B.~Tallman, A.~Boone, A.~Vijayakumar, F.~Lopez, S.~Apata, J.~Martinez et~al.,
  \emph{{Potential for definitive discovery of a 70 GeV dark matter WIMP with
  only second-order gauge couplings}},
  \href{https://arxiv.org/abs/2210.15019}{{\ttfamily 2210.15019}}.

\bibitem{Kang:2022zqv}
S.~Kang, A.~Kar and S.~Scopel, \emph{{Halo-independent bounds on the
  non-relativistic effective theory of WIMP-nucleon scattering from direct
  detection and neutrino observations}},
  \href{https://arxiv.org/abs/2212.05774}{{\ttfamily 2212.05774}}.

\bibitem{Dutta:2022wdi}
K.~Dutta, A.~Ghosh, A.~Kar and B.~Mukhopadhyaya, \emph{{MeV to multi-TeV
  thermal WIMPs are all observationally allowed}},
  \href{https://arxiv.org/abs/2212.09795}{{\ttfamily 2212.09795}}.

\bibitem{Akerib:2016vxi}
{\scshape LUX} collaboration, D.~S. Akerib et~al., \emph{{Results from a search
  for dark matter in the complete LUX exposure}},
  \href{http://dx.doi.org/10.1103/PhysRevLett.118.021303}{\emph{Phys. Rev.
  Lett.} {\bfseries 118} (2017) 021303},
  [\href{https://arxiv.org/abs/1608.07648}{{\ttfamily 1608.07648}}].

\bibitem{Zhang:2018xdp}
{\scshape PandaX} collaboration, H.~Zhang et~al., \emph{{Dark matter direct
  search sensitivity of the PandaX-4T experiment}},
  \href{http://dx.doi.org/10.1007/s11433-018-9259-0}{\emph{Sci. China Phys.
  Mech. Astron.} {\bfseries 62} (2019) 31011},
  [\href{https://arxiv.org/abs/1806.02229}{{\ttfamily 1806.02229}}].

\bibitem{Aprile:2018dbl}
{\scshape XENON} collaboration, E.~Aprile et~al., \emph{{Dark Matter Search
  Results from a One Ton-Year Exposure of XENON1T}},
  \href{http://dx.doi.org/10.1103/PhysRevLett.121.111302}{\emph{Phys. Rev.
  Lett.} {\bfseries 121} (2018) 111302},
  [\href{https://arxiv.org/abs/1805.12562}{{\ttfamily 1805.12562}}].

\bibitem{MAGIC:2016xys}
{\scshape MAGIC, Fermi-LAT} collaboration, M.~L. Ahnen et~al., \emph{{Limits to
  Dark Matter Annihilation Cross-Section from a Combined Analysis of MAGIC and
  Fermi-LAT Observations of Dwarf Satellite Galaxies}},
  \href{http://dx.doi.org/10.1088/1475-7516/2016/02/039}{\emph{JCAP} {\bfseries
  02} (2016) 039}, [\href{https://arxiv.org/abs/1601.06590}{{\ttfamily
  1601.06590}}].

\bibitem{Chatrchyan:2012xdj}
{\scshape CMS} collaboration, S.~Chatrchyan et~al., \emph{{Observation of a New
  Boson at a Mass of 125 GeV with the CMS Experiment at the LHC}},
  \href{http://dx.doi.org/10.1016/j.physletb.2012.08.021}{\emph{Phys. Lett.}
  {\bfseries B716} (2012) 30--61},
  [\href{https://arxiv.org/abs/1207.7235}{{\ttfamily 1207.7235}}].

\bibitem{Aad:2012tfa}
{\scshape ATLAS} collaboration, G.~Aad et~al., \emph{{Observation of a new
  particle in the search for the Standard Model Higgs boson with the ATLAS
  detector at the LHC}},
  \href{http://dx.doi.org/10.1016/j.physletb.2012.08.020}{\emph{Phys. Lett.}
  {\bfseries B716} (2012) 1--29},
  [\href{https://arxiv.org/abs/1207.7214}{{\ttfamily 1207.7214}}].

\bibitem{Hall:2009bx}
L.~J. Hall, K.~Jedamzik, J.~March-Russell and S.~M. West, \emph{{Freeze-In
  Production of FIMP Dark Matter}},
  \href{http://dx.doi.org/10.1007/JHEP03(2010)080}{\emph{JHEP} {\bfseries 03}
  (2010) 080}, [\href{https://arxiv.org/abs/0911.1120}{{\ttfamily 0911.1120}}].

\bibitem{Konar:2021oye}
P.~Konar, R.~Roshan and S.~Show, \emph{{Freeze-in dark matter through forbidden
  channel in U(1)B-L}},
  \href{http://dx.doi.org/10.1088/1475-7516/2022/03/021}{\emph{JCAP} {\bfseries
  03} (2022) 021}, [\href{https://arxiv.org/abs/2110.14411}{{\ttfamily
  2110.14411}}].

\bibitem{Ghosh:2021wrk}
P.~Ghosh, P.~Konar, A.~K. Saha and S.~Show, \emph{{Self-interacting freeze-in
  dark matter in a singlet doublet scenario}},
  \href{http://dx.doi.org/10.1088/1475-7516/2022/10/017}{\emph{JCAP} {\bfseries
  10} (2022) 017}, [\href{https://arxiv.org/abs/2112.09057}{{\ttfamily
  2112.09057}}].

\bibitem{Chakrabarty:2022bcn}
N.~Chakrabarty, P.~Konar, R.~Roshan~and and S.~Show, \emph{{Thermally corrected
  masses and freeze-in dark matter: a case study}},
  \href{https://arxiv.org/abs/2206.02233}{{\ttfamily 2206.02233}}.

\bibitem{DEramo:2017gpl}
F.~D'Eramo, N.~Fernandez and S.~Profumo, \emph{{When the Universe Expands Too
  Fast: Relentless Dark Matter}},
  \href{http://dx.doi.org/10.1088/1475-7516/2017/05/012}{\emph{JCAP} {\bfseries
  05} (2017) 012}, [\href{https://arxiv.org/abs/1703.04793}{{\ttfamily
  1703.04793}}].

\bibitem{DEramo:2017ecx}
F.~D'Eramo, N.~Fernandez and S.~Profumo, \emph{{Dark Matter Freeze-in
  Production in Fast-Expanding Universes}},
  \href{http://dx.doi.org/10.1088/1475-7516/2018/02/046}{\emph{JCAP} {\bfseries
  02} (2018) 046}, [\href{https://arxiv.org/abs/1712.07453}{{\ttfamily
  1712.07453}}].

\bibitem{Calibbi:2018fqf}
L.~Calibbi, L.~Lopez-Honorez, S.~Lowette and A.~Mariotti,
  \emph{{Singlet-Doublet Dark Matter Freeze-in: LHC displaced signatures versus
  cosmology}}, \href{http://dx.doi.org/10.1007/JHEP09(2018)037}{\emph{JHEP}
  {\bfseries 09} (2018) 037},
  [\href{https://arxiv.org/abs/1805.04423}{{\ttfamily 1805.04423}}].

\bibitem{No:2019gvl}
J.~M. No, P.~Tunney and B.~Zaldivar, \emph{{Probing Dark Matter freeze-in with
  long-lived particle signatures: MATHUSLA, HL-LHC and FCC-hh}},
  \href{http://dx.doi.org/10.1007/JHEP03(2020)022}{\emph{JHEP} {\bfseries 03}
  (2020) 022}, [\href{https://arxiv.org/abs/1908.11387}{{\ttfamily
  1908.11387}}].

\bibitem{Viel:2013fqw}
M.~Viel, G.~D. Becker, J.~S. Bolton and M.~G. Haehnelt, \emph{{Warm dark matter
  as a solution to the small scale crisis: New constraints from high redshift
  Lyman-\ensuremath{\alpha} forest data}},
  \href{http://dx.doi.org/10.1103/PhysRevD.88.043502}{\emph{Phys. Rev. D}
  {\bfseries 88} (2013) 043502},
  [\href{https://arxiv.org/abs/1306.2314}{{\ttfamily 1306.2314}}].

\bibitem{Yeche:2017upn}
C.~Y\`eche, N.~Palanque-Delabrouille, J.~Baur and H.~du~Mas~des Bourboux,
  \emph{{Constraints on neutrino masses from Lyman-alpha forest power spectrum
  with BOSS and XQ-100}},
  \href{http://dx.doi.org/10.1088/1475-7516/2017/06/047}{\emph{JCAP} {\bfseries
  06} (2017) 047}, [\href{https://arxiv.org/abs/1702.03314}{{\ttfamily
  1702.03314}}].

\bibitem{Irsic:2017ixq}
V.~Ir\v{s}i\v{c} et~al., \emph{{New Constraints on the free-streaming of warm
  dark matter from intermediate and small scale Lyman-$\alpha$ forest data}},
  \href{http://dx.doi.org/10.1103/PhysRevD.96.023522}{\emph{Phys. Rev. D}
  {\bfseries 96} (2017) 023522},
  [\href{https://arxiv.org/abs/1702.01764}{{\ttfamily 1702.01764}}].

\bibitem{Bode:2000gq}
P.~Bode, J.~P. Ostriker and N.~Turok, \emph{{Halo formation in warm dark matter
  models}}, \href{http://dx.doi.org/10.1086/321541}{\emph{Astrophys. J.}
  {\bfseries 556} (2001) 93--107},
  [\href{https://arxiv.org/abs/astro-ph/0010389}{{\ttfamily
  astro-ph/0010389}}].

\bibitem{Lovell_2012}
M.~R. Lovell, V.~Eke, C.~S. Frenk, L.~Gao, A.~Jenkins, T.~Theuns et~al.,
  \emph{The haloes of bright satellite galaxies in a warm dark matter
  universe},
  \href{http://dx.doi.org/10.1111/j.1365-2966.2011.20200.x}{\emph{Monthly
  Notices of the Royal Astronomical Society} {\bfseries 420} (jan, 2012)
  2318--2324}.

\bibitem{Yaguna:2015mva}
C.~E. Yaguna, \emph{{Singlet-Doublet Dirac Dark Matter}},
  \href{http://dx.doi.org/10.1103/PhysRevD.92.115002}{\emph{Phys. Rev.}
  {\bfseries D92} (2015) 115002},
  [\href{https://arxiv.org/abs/1510.06151}{{\ttfamily 1510.06151}}].

\bibitem{Fiaschi:2018rky}
J.~Fiaschi, M.~Klasen and S.~May, \emph{{Singlet-doublet fermion and triplet
  scalar dark matter with radiative neutrino masses}},
  \href{http://dx.doi.org/10.1007/JHEP05(2019)015}{\emph{JHEP} {\bfseries 05}
  (2019) 015}, [\href{https://arxiv.org/abs/1812.11133}{{\ttfamily
  1812.11133}}].

\bibitem{Restrepo:2019soi}
D.~Restrepo, A.~Rivera and W.~Tangarife, \emph{{Singlet-Doublet Dirac Dark
  Matter and Neutrino Masses}},
  \href{http://dx.doi.org/10.1103/PhysRevD.100.035029}{\emph{Phys. Rev.}
  {\bfseries D100} (2019) 035029},
  [\href{https://arxiv.org/abs/1906.09685}{{\ttfamily 1906.09685}}].

\bibitem{Arcadi:2018pfo}
G.~Arcadi, \emph{{2HDM portal for Singlet-Doublet Dark Matter}},
  \href{http://dx.doi.org/10.1140/epjc/s10052-018-6327-6}{\emph{Eur. Phys. J.}
  {\bfseries C78} (2018) 864},
  [\href{https://arxiv.org/abs/1804.04930}{{\ttfamily 1804.04930}}].

\bibitem{Esch:2018ccs}
S.~Esch, M.~Klasen and C.~E. Yaguna, \emph{{A singlet doublet dark matter model
  with radiative neutrino masses}},
  \href{http://dx.doi.org/10.1007/JHEP10(2018)055}{\emph{JHEP} {\bfseries 10}
  (2018) 055}, [\href{https://arxiv.org/abs/1804.03384}{{\ttfamily
  1804.03384}}].

\bibitem{Maru:2017pwl}
N.~Maru, N.~Okada and S.~Okada, \emph{{Fermionic Minimal Dark Matter in 5D
  Gauge-Higgs Unification}},
  \href{http://dx.doi.org/10.1103/PhysRevD.96.115023}{\emph{Phys. Rev.}
  {\bfseries D96} (2017) 115023},
  [\href{https://arxiv.org/abs/1801.00686}{{\ttfamily 1801.00686}}].

\bibitem{Maru:2017otg}
N.~Maru, T.~Miyaji, N.~Okada and S.~Okada, \emph{{Fermion Dark Matter in
  Gauge-Higgs Unification}},
  \href{http://dx.doi.org/10.1007/JHEP07(2017)048}{\emph{JHEP} {\bfseries 07}
  (2017) 048}, [\href{https://arxiv.org/abs/1704.04621}{{\ttfamily
  1704.04621}}].

\bibitem{Xiang:2017yfs}
Q.-F. Xiang, X.-J. Bi, P.-F. Yin and Z.-H. Yu, \emph{{Exploring Fermionic Dark
  Matter via Higgs Boson Precision Measurements at the Circular Electron
  Positron Collider}},
  \href{http://dx.doi.org/10.1103/PhysRevD.97.055004}{\emph{Phys. Rev.}
  {\bfseries D97} (2018) 055004},
  [\href{https://arxiv.org/abs/1707.03094}{{\ttfamily 1707.03094}}].

\bibitem{Abe:2017glm}
T.~Abe, \emph{{Effect of CP violation in the singlet-doublet dark matter
  model}}, \href{http://dx.doi.org/10.1016/j.physletb.2017.05.048}{\emph{Phys.
  Lett.} {\bfseries B771} (2017) 125--130},
  [\href{https://arxiv.org/abs/1702.07236}{{\ttfamily 1702.07236}}].

\bibitem{Banerjee:2016hsk}
S.~Banerjee, S.~Matsumoto, K.~Mukaida and Y.-L.~S. Tsai, \emph{{WIMP Dark
  Matter in a Well-Tempered Regime: A case study on Singlet-Doublets Fermionic
  WIMP}}, \href{http://dx.doi.org/10.1007/JHEP11(2016)070}{\emph{JHEP}
  {\bfseries 11} (2016) 070},
  [\href{https://arxiv.org/abs/1603.07387}{{\ttfamily 1603.07387}}].

\bibitem{Horiuchi:2016tqw}
S.~Horiuchi, O.~Macias, D.~Restrepo, A.~Rivera, O.~Zapata and H.~Silverwood,
  \emph{{The Fermi-LAT gamma-ray excess at the Galactic Center in the
  singlet-doublet fermion dark matter model}},
  \href{http://dx.doi.org/10.1088/1475-7516/2016/03/048}{\emph{JCAP} {\bfseries
  1603} (2016) 048}, [\href{https://arxiv.org/abs/1602.04788}{{\ttfamily
  1602.04788}}].

\bibitem{Calibbi:2015nha}
L.~Calibbi, A.~Mariotti and P.~Tziveloglou, \emph{{Singlet-Doublet Model: Dark
  matter searches and LHC constraints}},
  \href{http://dx.doi.org/10.1007/JHEP10(2015)116}{\emph{JHEP} {\bfseries 10}
  (2015) 116}, [\href{https://arxiv.org/abs/1505.03867}{{\ttfamily
  1505.03867}}].

\bibitem{Cheung:2013dua}
C.~Cheung and D.~Sanford, \emph{{Simplified Models of Mixed Dark Matter}},
  \href{http://dx.doi.org/10.1088/1475-7516/2014/02/011}{\emph{JCAP} {\bfseries
  1402} (2014) 011}, [\href{https://arxiv.org/abs/1311.5896}{{\ttfamily
  1311.5896}}].

\bibitem{Cohen:2011ec}
T.~Cohen, J.~Kearney, A.~Pierce and D.~Tucker-Smith, \emph{{Singlet-Doublet
  Dark Matter}},
  \href{http://dx.doi.org/10.1103/PhysRevD.85.075003}{\emph{Phys. Rev.}
  {\bfseries D85} (2012) 075003},
  [\href{https://arxiv.org/abs/1109.2604}{{\ttfamily 1109.2604}}].

\bibitem{Enberg:2007rp}
R.~Enberg, P.~J. Fox, L.~J. Hall, A.~Y. Papaioannou and M.~Papucci, \emph{{LHC
  and dark matter signals of improved naturalness}},
  \href{http://dx.doi.org/10.1088/1126-6708/2007/11/014}{\emph{JHEP} {\bfseries
  11} (2007) 014}, [\href{https://arxiv.org/abs/0706.0918}{{\ttfamily
  0706.0918}}].

\bibitem{DEramo:2007anh}
F.~D'Eramo, \emph{{Dark matter and Higgs boson physics}},
  \href{http://dx.doi.org/10.1103/PhysRevD.76.083522}{\emph{Phys. Rev.}
  {\bfseries D76} (2007) 083522},
  [\href{https://arxiv.org/abs/0705.4493}{{\ttfamily 0705.4493}}].

\bibitem{Barman:2019aku}
B.~Barman, D.~Borah, P.~Ghosh and A.~K. Saha, \emph{{Flavoured gauge extension
  of singlet-doublet fermionic dark matter: neutrino mass, high scale validity
  and collider signatures}},
  \href{http://dx.doi.org/10.1007/JHEP10(2019)275}{\emph{JHEP} {\bfseries 10}
  (2019) 275}, [\href{https://arxiv.org/abs/1907.10071}{{\ttfamily
  1907.10071}}].

\bibitem{DuttaBanik:2018emv}
A.~Dutta~Banik, A.~K. Saha and A.~Sil, \emph{{Scalar assisted singlet doublet
  fermion dark matter model and electroweak vacuum stability}},
  \href{http://dx.doi.org/10.1103/PhysRevD.98.075013}{\emph{Phys. Rev.}
  {\bfseries D98} (2018) 075013},
  [\href{https://arxiv.org/abs/1806.08080}{{\ttfamily 1806.08080}}].

\bibitem{Barman:2019tuo}
B.~Barman, S.~Bhattacharya, P.~Ghosh, S.~Kadam and N.~Sahu, \emph{{Fermion Dark
  Matter with Scalar Triplet at Direct and Collider Searches}},
  \href{http://dx.doi.org/10.1103/PhysRevD.100.015027}{\emph{Phys. Rev.}
  {\bfseries D100} (2019) 015027},
  [\href{https://arxiv.org/abs/1902.01217}{{\ttfamily 1902.01217}}].

\bibitem{Bhattacharya:2018fus}
S.~Bhattacharya, P.~Ghosh, N.~Sahoo and N.~Sahu, \emph{{Mini Review on
  Vector-Like Leptonic Dark Matter, Neutrino Mass, and Collider Signatures}},
  \href{http://dx.doi.org/10.3389/fphy.2019.00080}{\emph{Front.in Phys.}
  {\bfseries 7} (2019) 80}, [\href{https://arxiv.org/abs/1812.06505}{{\ttfamily
  1812.06505}}].

\bibitem{Bhattacharya:2015qpa}
S.~Bhattacharya, N.~Sahoo and N.~Sahu, \emph{{Minimal vectorlike leptonic dark
  matter and signatures at the LHC}},
  \href{http://dx.doi.org/10.1103/PhysRevD.93.115040}{\emph{Phys. Rev.}
  {\bfseries D93} (2016) 115040},
  [\href{https://arxiv.org/abs/1510.02760}{{\ttfamily 1510.02760}}].

\bibitem{Bhattacharya:2017sml}
S.~Bhattacharya, N.~Sahoo and N.~Sahu, \emph{{Singlet-Doublet Fermionic Dark
  Matter, Neutrino Mass and Collider Signatures}},
  \href{http://dx.doi.org/10.1103/PhysRevD.96.035010}{\emph{Phys. Rev.}
  {\bfseries D96} (2017) 035010},
  [\href{https://arxiv.org/abs/1704.03417}{{\ttfamily 1704.03417}}].

\bibitem{Restrepo:2015ura}
D.~Restrepo, A.~Rivera, M.~Sánchez-Peláez, O.~Zapata and W.~Tangarife,
  \emph{{Radiative Neutrino Masses in the Singlet-Doublet Fermion Dark Matter
  Model with Scalar Singlets}},
  \href{http://dx.doi.org/10.1103/PhysRevD.92.013005}{\emph{Phys. Rev.}
  {\bfseries D92} (2015) 013005},
  [\href{https://arxiv.org/abs/1504.07892}{{\ttfamily 1504.07892}}].

\bibitem{Freitas:2015hsa}
A.~Freitas, S.~Westhoff and J.~Zupan, \emph{{Integrating in the Higgs Portal to
  Fermion Dark Matter}},
  \href{http://dx.doi.org/10.1007/JHEP09(2015)015}{\emph{JHEP} {\bfseries 09}
  (2015) 015}, [\href{https://arxiv.org/abs/1506.04149}{{\ttfamily
  1506.04149}}].

\bibitem{Cynolter:2015sua}
G.~Cynolter, J.~Kovács and E.~Lendvai, \emph{{Doublet–singlet model and
  unitarity}}, \href{http://dx.doi.org/10.1142/S0217732316500139}{\emph{Mod.
  Phys. Lett.} {\bfseries A31} (2016) 1650013},
  [\href{https://arxiv.org/abs/1509.05323}{{\ttfamily 1509.05323}}].

\bibitem{Bhattacharya:2016lts}
S.~Bhattacharya, B.~Karmakar, N.~Sahu and A.~Sil, \emph{{Unifying the flavor
  origin of dark matter with leptonic nonzero $\theta_{13}$}},
  \href{http://dx.doi.org/10.1103/PhysRevD.93.115041}{\emph{Phys. Rev.}
  {\bfseries D93} (2016) 115041},
  [\href{https://arxiv.org/abs/1603.04776}{{\ttfamily 1603.04776}}].

\bibitem{Bhattacharya:2016rqj}
S.~Bhattacharya, B.~Karmakar, N.~Sahu and A.~Sil, \emph{{Flavor origin of dark
  matter and its relation with leptonic nonzero $\theta_{13}$ and Dirac CP
  phase $\delta$}},
  \href{http://dx.doi.org/10.1007/JHEP05(2017)068}{\emph{JHEP} {\bfseries 05}
  (2017) 068}, [\href{https://arxiv.org/abs/1611.07419}{{\ttfamily
  1611.07419}}].

\bibitem{Wang:2018lhk}
J.-W. Wang, X.-J. Bi, P.-F. Yin and Z.-H. Yu, \emph{{Impact of Fermionic
  Electroweak Multiplet Dark Matter on Vacuum Stability with One-loop
  Matching}}, \href{http://dx.doi.org/10.1103/PhysRevD.99.055009}{\emph{Phys.
  Rev.} {\bfseries D99} (2019) 055009},
  [\href{https://arxiv.org/abs/1811.08743}{{\ttfamily 1811.08743}}].

\bibitem{Abe:2019wku}
T.~Abe and R.~Sato, \emph{{Current status and future prospects of the
  singlet-doublet dark matter model with CP-violation}},
  \href{http://dx.doi.org/10.1103/PhysRevD.99.035012}{\emph{Phys. Rev.}
  {\bfseries D99} (2019) 035012},
  [\href{https://arxiv.org/abs/1901.02278}{{\ttfamily 1901.02278}}].

\bibitem{Barman:2019oda}
B.~Barman, A.~Dutta~Banik and A.~Paul, \emph{{Singlet-doublet fermionic dark
  matter and gravitational waves in a two-Higgs-doublet extension of the
  Standard Model}},
  \href{http://dx.doi.org/10.1103/PhysRevD.101.055028}{\emph{Phys. Rev. D}
  {\bfseries 101} (2020) 055028},
  [\href{https://arxiv.org/abs/1912.12899}{{\ttfamily 1912.12899}}].

\bibitem{Elor:2021swj}
G.~Elor, R.~McGehee and A.~Pierce, \emph{{Maximizing Direct Detection with
  HYPER Dark Matter}},  \href{https://arxiv.org/abs/2112.03920}{{\ttfamily
  2112.03920}}.

\bibitem{Bhattiprolu:2022sdd}
P.~N. Bhattiprolu, G.~Elor, R.~McGehee and A.~Pierce, \emph{{Freezing-in
  hadrophilic dark matter at low reheating temperatures}},
  \href{https://arxiv.org/abs/2210.15653}{{\ttfamily 2210.15653}}.

\bibitem{Thomas:1998wy}
S.~D. Thomas and J.~D. Wells, \emph{{Phenomenology of Massive Vectorlike
  Doublet Leptons}},
  \href{http://dx.doi.org/10.1103/PhysRevLett.81.34}{\emph{Phys. Rev. Lett.}
  {\bfseries 81} (1998) 34--37},
  [\href{https://arxiv.org/abs/hep-ph/9804359}{{\ttfamily hep-ph/9804359}}].

\bibitem{Cirelli:2009uv}
M.~Cirelli and A.~Strumia, \emph{{Minimal Dark Matter: Model and results}},
  \href{http://dx.doi.org/10.1088/1367-2630/11/10/105005}{\emph{New J. Phys.}
  {\bfseries 11} (2009) 105005},
  [\href{https://arxiv.org/abs/0903.3381}{{\ttfamily 0903.3381}}].

\bibitem{Haque:2023yra}
M.~R. Haque, D.~Maity and R.~Mondal, \emph{{WIMPs, FIMPs, and Inflaton
  phenomenology via reheating, CMB and $\Delta N_{eff}$}},
  \href{https://arxiv.org/abs/2301.01641}{{\ttfamily 2301.01641}}.

\bibitem{Alloul:2013bka}
A.~Alloul, N.~D. Christensen, C.~Degrande, C.~Duhr and B.~Fuks,
  \emph{{FeynRules 2.0 - A complete toolbox for tree-level phenomenology}},
  \href{http://dx.doi.org/10.1016/j.cpc.2014.04.012}{\emph{Comput. Phys.
  Commun.} {\bfseries 185} (2014) 2250--2300},
  [\href{https://arxiv.org/abs/1310.1921}{{\ttfamily 1310.1921}}].

\bibitem{Degrande:2011ua}
C.~Degrande, C.~Duhr, B.~Fuks, D.~Grellscheid, O.~Mattelaer and T.~Reiter,
  \emph{{UFO - The Universal FeynRules Output}},
  \href{http://dx.doi.org/10.1016/j.cpc.2012.01.022}{\emph{Comput. Phys.
  Commun.} {\bfseries 183} (2012) 1201--1214},
  [\href{https://arxiv.org/abs/1108.2040}{{\ttfamily 1108.2040}}].

\bibitem{Alwall:2014hca}
J.~Alwall, R.~Frederix, S.~Frixione, V.~Hirschi, F.~Maltoni, O.~Mattelaer
  et~al., \emph{{The automated computation of tree-level and next-to-leading
  order differential cross sections, and their matching to parton shower
  simulations}}, \href{http://dx.doi.org/10.1007/JHEP07(2014)079}{\emph{JHEP}
  {\bfseries 07} (2014) 079},
  [\href{https://arxiv.org/abs/1405.0301}{{\ttfamily 1405.0301}}].

\bibitem{Bierlich:2022pfr}
C.~Bierlich et~al., \emph{{A comprehensive guide to the physics and usage of
  PYTHIA 8.3}},  \href{https://arxiv.org/abs/2203.11601}{{\ttfamily
  2203.11601}}.

\bibitem{Mangano:2006rw}
M.~L. Mangano, M.~Moretti, F.~Piccinini and M.~Treccani, \emph{{Matching matrix
  elements and shower evolution for top-quark production in hadronic
  collisions}},
  \href{http://dx.doi.org/10.1088/1126-6708/2007/01/013}{\emph{JHEP} {\bfseries
  01} (2007) 013}, [\href{https://arxiv.org/abs/hep-ph/0611129}{{\ttfamily
  hep-ph/0611129}}].

\bibitem{deFavereau:2013fsa}
{\scshape DELPHES 3} collaboration, J.~de~Favereau, C.~Delaere, P.~Demin,
  A.~Giammanco, V.~Lema\^\i{}tre, A.~Mertens et~al., \emph{{DELPHES 3, A
  modular framework for fast simulation of a generic collider experiment}},
  \href{http://dx.doi.org/10.1007/JHEP02(2014)057}{\emph{JHEP} {\bfseries 02}
  (2014) 057}, [\href{https://arxiv.org/abs/1307.6346}{{\ttfamily 1307.6346}}].

\bibitem{Cacciari:2011ma}
M.~Cacciari, G.~P. Salam and G.~Soyez, \emph{{FastJet User Manual}},
  \href{http://dx.doi.org/10.1140/epjc/s10052-012-1896-2}{\emph{Eur. Phys. J.
  C} {\bfseries 72} (2012) 1896},
  [\href{https://arxiv.org/abs/1111.6097}{{\ttfamily 1111.6097}}].

\bibitem{Cacciari:2008gp}
M.~Cacciari, G.~P. Salam and G.~Soyez, \emph{{The anti-$k_t$ jet clustering
  algorithm}},
  \href{http://dx.doi.org/10.1088/1126-6708/2008/04/063}{\emph{JHEP} {\bfseries
  04} (2008) 063}, [\href{https://arxiv.org/abs/0802.1189}{{\ttfamily
  0802.1189}}].

\bibitem{Dokshitzer:1997in}
Y.~L. Dokshitzer, G.~D. Leder, S.~Moretti and B.~R. Webber, \emph{{Better jet
  clustering algorithms}},
  \href{http://dx.doi.org/10.1088/1126-6708/1997/08/001}{\emph{JHEP} {\bfseries
  08} (1997) 001}, [\href{https://arxiv.org/abs/hep-ph/9707323}{{\ttfamily
  hep-ph/9707323}}].

\bibitem{CMS:2009lxa}
{\scshape CMS} collaboration, \emph{{A Cambridge-Aachen (C-A) based Jet
  Algorithm for boosted top-jet tagging}}, .

\bibitem{CMS:2017yfk}
{\scshape CMS} collaboration, A.~M. Sirunyan et~al., \emph{{Particle-flow
  reconstruction and global event description with the CMS detector}},
  \href{http://dx.doi.org/10.1088/1748-0221/12/10/P10003}{\emph{JINST}
  {\bfseries 12} (2017) P10003},
  [\href{https://arxiv.org/abs/1706.04965}{{\ttfamily 1706.04965}}].

\bibitem{Bertolini:2014bba}
D.~Bertolini, P.~Harris, M.~Low and N.~Tran, \emph{{Pileup Per Particle
  Identification}},
  \href{http://dx.doi.org/10.1007/JHEP10(2014)059}{\emph{JHEP} {\bfseries 10}
  (2014) 059}, [\href{https://arxiv.org/abs/1407.6013}{{\ttfamily 1407.6013}}].

\bibitem{Cacciari:2014gra}
M.~Cacciari, G.~P. Salam and G.~Soyez, \emph{{SoftKiller, a particle-level
  pileup removal method}},
  \href{http://dx.doi.org/10.1140/epjc/s10052-015-3267-2}{\emph{Eur. Phys. J.
  C} {\bfseries 75} (2015) 59},
  [\href{https://arxiv.org/abs/1407.0408}{{\ttfamily 1407.0408}}].

\bibitem{CMS:2020ebo}
{\scshape CMS} collaboration, A.~M. Sirunyan et~al., \emph{{Pileup mitigation
  at CMS in 13 TeV data}},
  \href{http://dx.doi.org/10.1088/1748-0221/15/09/P09018}{\emph{JINST}
  {\bfseries 15} (2020) P09018},
  [\href{https://arxiv.org/abs/2003.00503}{{\ttfamily 2003.00503}}].

\bibitem{BRUN199781}
R.~Brun and F.~Rademakers, \emph{Root — an object oriented data analysis
  framework},
  \href{http://dx.doi.org/https://doi.org/10.1016/S0168-9002(97)00048-X}{\emph{Nuclear
  Instruments and Methods in Physics Research Section A: Accelerators,
  Spectrometers, Detectors and Associated Equipment} {\bfseries 389} (1997)
  81--86}.

\bibitem{Hocker:2007ht}
A.~Hocker et~al., \emph{{TMVA - Toolkit for Multivariate Data Analysis}},
  \href{https://arxiv.org/abs/physics/0703039}{{\ttfamily physics/0703039}}.

\bibitem{prune}
S.~D. Ellis, C.~K. Vermilion and J.~R. Walsh, \emph{Recombination algorithms
  and jet substructure: Pruning as a tool for heavy particle searches},
  \href{http://dx.doi.org/10.1103/PhysRevD.81.094023}{\emph{Phys. Rev. D}
  {\bfseries 81} (May, 2010) 094023}.

\bibitem{Larkoski:2014wba}
A.~J. Larkoski, S.~Marzani, G.~Soyez and J.~Thaler, \emph{{Soft Drop}},
  \href{http://dx.doi.org/10.1007/JHEP05(2014)146}{\emph{JHEP} {\bfseries 05}
  (2014) 146}, [\href{https://arxiv.org/abs/1402.2657}{{\ttfamily 1402.2657}}].

\bibitem{Larkoski:2013eya}
A.~J. Larkoski, G.~P. Salam and J.~Thaler, \emph{{Energy Correlation Functions
  for Jet Substructure}},
  \href{http://dx.doi.org/10.1007/JHEP06(2013)108}{\emph{JHEP} {\bfseries 06}
  (2013) 108}, [\href{https://arxiv.org/abs/1305.0007}{{\ttfamily 1305.0007}}].

\bibitem{Thaler:2010tr}
J.~Thaler and K.~Van~Tilburg, \emph{{Identifying Boosted Objects with
  N-subjettiness}},
  \href{http://dx.doi.org/10.1007/JHEP03(2011)015}{\emph{JHEP} {\bfseries 03}
  (2011) 015}, [\href{https://arxiv.org/abs/1011.2268}{{\ttfamily 1011.2268}}].

\bibitem{CMS:2018szt}
{\scshape CMS} collaboration, A.~M. Sirunyan et~al., \emph{{Combined search for
  electroweak production of charginos and neutralinos in proton-proton
  collisions at $\sqrt{s} =$ 13 TeV}},
  \href{http://dx.doi.org/10.1007/JHEP03(2018)160}{\emph{JHEP} {\bfseries 03}
  (2018) 160}, [\href{https://arxiv.org/abs/1801.03957}{{\ttfamily
  1801.03957}}].

\end{thebibliography}\endgroup
\end{document}